\documentclass{article}
\usepackage{arxiv}

\usepackage[utf8]{inputenc} 
\usepackage[T1]{fontenc}    
\usepackage[colorlinks]{hyperref}       
\hypersetup{
  colorlinks   = true, 
  urlcolor     = red, 
  linkcolor    = blue,
  citecolor   = blue
}
\usepackage{url}            
\usepackage{booktabs}      
\usepackage{amsfonts}       
\usepackage{nicefrac}       
\usepackage{microtype}  
\usepackage{graphicx}
\usepackage{caption}
\usepackage[round]{natbib}
\usepackage{appendix}
\usepackage{amsthm,bm,amsmath,amssymb,epsfig,multicol}
\usepackage{graphics,rotating, xcolor}
\usepackage{array,multirow}
\usepackage{typearea}
\usepackage{pdflscape}
\usepackage{diagbox, makecell}
\usepackage{collcell,datatool}
\usepackage{subfigure}
\usepackage{hhline}
\usepackage{etoolbox}

\newcommand{\edr}{e.d.r }
\newcommand{\bbeta}{\bm{\beta}}
\newcommand{\bX}{\bm{\beta}^\top \mathbf{X}}
\newcommand{\bXone}{\bm{\beta}_1^\top \mathbf{X}}
\newcommand{\bXk}{\bm{\beta}_K^\top \mathbf{X}}
\newcommand{\bols}{\bm{b}_{ols}}
\newcommand{\Sb}{\mathcal{S_B }}

\newcommand{\BX}{\boldsymbol{\mathbf{B}}^\top \mathbf{X}}
\newcommand{\X}{\mathbf{X}}
\newcommand{\Z}{\mathbf{Z}}
\newcommand{\E}{\text{E}}
\newcommand{\C}{\bm{\Sigma}\mathbf{B}\left(\mathbf{B}^\top\bm{\Sigma}\mathbf{B}\right)^{-1}\mathbf{B}^\top}

\newcommand{\Si}{\bm{\Sigma}}
\newcommand{\B}{\bm{B}}
\newcommand{\bh}{\bm{b}_h}
\newcommand{\m}{\bm{\mu}}
\newcommand{\e}{\varepsilon}
\newcommand{\Ge}{G_\e}

\newcommand{\g}{\bm{\gamma}}
\newcommand{\ShInv}{\Si_h^{-1}}
\newcommand{\Sxyh}{\Si_{xy,h}}

\newcommand{\mh}{\bm{\mu}_h}
\newcommand{\x}{\bm{x}}

\newtheorem{thm}{Theorem}

\newtheorem{lem}{Lemma}

\theoremstyle{definition}

\newtheorem{remark}{Remark}
\newtheorem{condition}{Condition}

\newtheorem{assumption}{Assumption}

\patchcmd{\quote}{\rightmargin}{\leftmargin 0em \rightmargin}{}{}
\newcounter{modelcounter}
\newenvironment{model}{\begin{quote}%
    \refstepcounter{modelcounter}%
  \textbf{ Model \arabic{modelcounter}}%
}{%
\end{quote}%
}

\title{Slice weighted average regression}

\date{September 9, 2022}	

\author{{Marina Masioti \thanks{Corresponding author}} \\
	Department of Mathematical and Physical Sciences\\
	La Trobe University\\
	Melbourne, VIC 3086, Australia\\
	\texttt{mmasioti@students.latrobe.edu.au} \\
	\And
	Joshua Davies \\
	SAS Institute \\
	Melbourne, VIC 3004, Australia \\
	\texttt{josh.davies11@gmail.com} \\
	\And
	Amanda Shaker \\
	Department of Mathematical and Physical Sciences\\
	La Trobe University\\
	Melbourne, VIC 3086, Australia \\
	\texttt{A.Shaker@latrobe.edu.au} \\
	\And
	\href{https://scholars.latrobe.edu.au/lprendergast}{\includegraphics[scale=0.06]{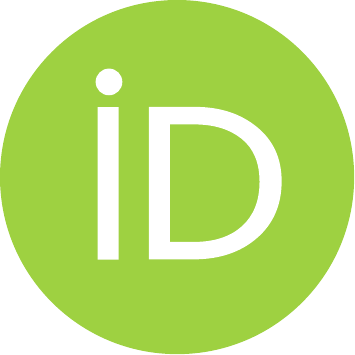}\hspace{1mm}Luke A. Prendergast}\\
	Department of Mathematical and Physical Sciences\\
	La Trobe University\\
	Melbourne, VIC 3086, Australia\\
	\texttt{luke.prendergast@latrobe.edu.au} \\
}

\renewcommand{\shorttitle}{\textit{SWAR}}

\hypersetup{
pdftitle={Slice Weighted Average Regression},
pdfsubject={},
pdfauthor={Marina Masioti, Joshua Davies, Luke A. Prendergast, Amanda Shaker },
pdfkeywords={Dimension Reduction, Influence Functions, Ordinary Least Squares, Sliced Inverse Regression},
}

\begin{document}

\maketitle

\begin{abstract}
It has previously been shown that ordinary least squares can be used to estimate the coefficients of the single-index model under only mild conditions.  However, the estimator is non-robust leading to poor estimates for some models.  In this paper we propose a new sliced least-squares estimator that utilizes ideas from Sliced Inverse Regression. Slices with problematic observations that contribute to high variability in the estimator can easily be down-weighted to robustify the procedure. The estimator is simple to implement and can result in vast improvements for some models when compared to the usual least-squares approach. While the estimator was initially conceived with the single-index model in mind, we also show that multiple directions can be obtained, therefore providing another notable advantage of using slicing with least squares.  Several simulation studies and a real data example are included, as well as some comparisons with some other recent methods.
\end{abstract}

\keywords{Dimension reduction, ordinary least squares, sliced inverse regression, influence functions.}

\section{Introduction}

Methodologies for dimension reduction have been immensely expanded in recent decades. It has become a prevalent topic due to the rapid advancements in computer technologies and the need for researchers to find complex structures in high-dimensional data sets. The `curse of dimensionality' \citep[e.g.][]{Bellman61} is commonly mentioned in dimension reduction settings since it describes the problem where as dimensionality gets higher, larger sample sizes are required to produce accurate estimates. Standard regression methods often assume simple regression models and can therefore fail at identifying more complicated relationships between a random univariate response variable, $Y \in \mathbb{R}$, and a random high-dimensional predictor variable $\X$. This is further complicated by our inability to sufficiently visualize high-dimensional data sets. The dimension reduction (DR) methods we consider here aim to reduce the dimensionality of the predictor vector by replacing it with one or more linear combinations of the predictor components.  When little regression information is lost, this allows for the visualization of the data in a lower-dimensional framework. \cite{Li91} described such a scenario by the following model where for,  $\X = [ X_1, \dots, X_p]^\top \in \mathbb{R}^p$, we have,  
\begin{equation}\label{eq:DRmodel}
    Y = f(\bXone,\dots, \bXk, \e)
\end{equation}
where $f$ is the unknown \textit{link function}, $\bm{\beta}_i$'s (for $i = 1,\dots, K)$ are the $p$-dimensional column vectors of coefficients and $\e$ is the error term independent of $\X$ with E$(\e)=0$. Dimension reduction is then achieved when we replace $\X$ with $\bXone,\dots, \bXk$ where $K < p$. Then, a plot of $Y$ versus the lower-dimensional projections, $\bXone,\dots, \bXk$, is referred to as a Sufficient Summary Plot \citep[SSP,][]{Cook98} and reveals the structure of $f$. In this setting, the aim of DR methods are to find a basis for the set $\mathcal{S_B} = \text{span}(\bm{\beta}_1, \dots, \bm{\beta}_K)$, referred to as the \textit{effective dimension reduction} (\edr) space whose elements are referred to as \edr directions. Throughout we assume that $\mathcal{S_B}$ denotes the Central Dimension Reduction Space \citep[CDRS,][]{Cook98}, defined as the intersection of all dimension reduction subspaces. In practice, $f$ is unknown and therefore we cannot uniquely identify $\bbeta_1, \dots, \bbeta_K$. However, any set of $p$-dimensional vectors, $\g_1, \dots, \g_K$, such that span$(\bbeta_1, \dots, \bbeta_K) = \text{span}(\g_1, \dots, \g_K)$, is sufficient. 

In recent decades there has been much interest in this form of dimension reduction.  Ordinary Least Squares (OLS) in the case of the single-index model (i.e. $K=1$) \citep{BR77, BR83, li1989} led to further advances in methods such as Sliced Inverse Regression \citep[SIR,][]{Li91}, Sliced Average Variance Estimation \citep[SAVE,][]{CookWeis91} and Principal Hessian Directions \citep[pHd,][]{Li92}, just to name a few. Each method has its strengths and limitations and works well under specific conditions and assumptions and with particular model types. For this reason, the combination of dimension reduction methods and statistical tools has been used numerous times to improve the estimation of the CDRS, see for example, \cite{MAVE}, \cite{YeWeiss}, \cite{CookYin2004}, \cite{ZHU2007}, \cite{CookPFC2007}, \cite{ForzaniPFC2008}, \cite{CookForzani2009}, \cite{PSVM},  \cite{PALS}. Based on the simplicity and good results that can be achieved by SIR by using slicing, our goal in this paper is to investigate the use of least squares within this slicing context.

We begin with a brief overview of OLS and SIR in Section \ref{sec:DRmethods}. The theory and implementation of the new method is presented in Section \ref{sec:SWAR}. Influence functions are derived in Section \ref{sec:IFs} and used to understand the behaviour of the method for various contamination structures. Slice weights used to down-weight influential slices are formulated in Section \ref{sec:IFweights}. The effectiveness of the method is highlighted via simulations and a real-data example, in Section \ref{sec:simulations} and \ref{sec:example}, respectively. We conclude with a discussion in Section \ref{sec:discussion}. Proofs and other supporting material are given in the Appendix.

\section{Dimension reduction methods}\label{sec:DRmethods}

As our main motivating methods, in this section we briefly consider OLS and SIR for dimension reduction.  Before we do, we provide the Linear Design Condition (LDC), defined in its general form for $K\geq 1$ by \cite{Li91}:
\begin{condition}[LDC]\label{LDC}
For any $\bm{b}\in \mathbb{R}^p$, there are some scalar constants $a_0, a_1, \dots, a_K$ such that, 
\[E\big( \bm{b}^\top \X | \bbeta_1^\top \X, \dots, \bbeta_K^\top \X \big) = a_0 + \sum_{i=1}^K a_i \bbeta_i^\top \X.\]
\end{condition} 
Condition \ref{LDC} is satisfied when $\X$ follows an elliptically symmetric distribution \citep{EATON1986}.  However, it has also been shown to often approximately hold when $p$ is large \citep{hall1993}. 

\subsection{Ordinary Least Squares}\label{subsec:OLS}
 
Ordinary least squares (OLS) is commonly used for estimating the unknown parameters in the multiple linear regression (MLR).   Using the notation above for the model in \eqref{eq:DRmodel}, the MLR model is $Y = \beta_0 + \bbeta_1^\top \X + \e$, where $\beta_0$ and $\bbeta_1$ are the intercept and slope vector respectively. However, the capabilities of OLS expand further than just MLR. Under the single-index model $K=1$, if $\X$ follows a multivariate normal distribution and an additive error is assumed, \cite{BR77, BR83} showed that the OLS slope, which we denote $\bols$, can be used to determine the direction of $\bbeta_1$ since $\bols=c\bbeta_1$ for some $c\in \mathbb{R}$, provided $c\neq 0$. \cite{li1989} extended the above for OLS to include milder distributional conditions for $\X$ (e.g. Condition \ref{LDC} with $K=1$) and a non-additive error for a single-index model of the form
\begin{equation}\label{eq:SIM}
Y = f(\bX, \e).
\end{equation}

Hence, OLS can be used much more extensively than for just the MLR model. However, OLS can fail for some model types where the slope vector is equal to zero.  This can occur when the model exhibits symmetric dependency (when the model is symmetric around the mean of $\bX$), but also for less obvious scenarios,  \cite[e.g.][Example 2.1]{garnham2013}. 

\subsection{Sliced Inverse Regression}\label{subsec:SIR}

\cite{Li91} introduced SIR to estimate the inverse regression curve, $\text{E}(\X | Y)$, which provides a notable advantage over the regression of $Y|\X$ when dealing with high-dimensional data. Let $\m=\text{E}(\X)$, $\Si = \text{Var}(\X)$ and the standardized regressor be defined as $\Z = \Si^{-1/2} (\X - \m)$.  \cite{Li91} showed that under Condition \ref{LDC}, $\E(\Z| Y)\in \Si^{1/2} \mathcal{S_B}$ for the model in \eqref{eq:DRmodel}. As a consequence, the eigenvectors of $\text{Var}\big[\text{E}(\Z|Y)\big]$ that correspond to the largest non-zero eigenvalues are elements of $\Si^{1/2} \mathcal{S_B}$ and a basis for $\Sb$ can be found by the re-standardization of the eigenvectors with respect to $\Si^{-1/2}$.

In practice, the estimation of the standardized inverse regression curve is performed by partitioning the data into several slices, based on the value of $Y$, which produces an approximating step function for E$(\bm{Z}|Y)$. It is common for a continuous response to choose $H$ equally proportioned slices, whereas in the discrete case, the slices are already naturally defined by the unique values of $Y$. Let $S_1, \dots, S_H$ denote the $H$ non-overlapping sub-ranges of $Y$ and $p_1, \dots, p_H$ the slice proportions, i.e. $p_h=P(Y\in S_h)$. The SIR matrix is then given by 
\begin{equation}
    \mathbf{V} = \Si^{-1/2} \displaystyle\sum_{h=1}^H p_h (\mh - \m) (\mh - \m)^\top \Si^{-1/2},
\end{equation}
where $\mh = \text{E}(\X|Y \in S_h)$ is the $h$th slice mean. Since $\m$ is a linear combination of the slice means, the maximum rank of $\mathbf{V}$ is $H-1$.  Under Condition \ref{LDC}, eigenvectors of $\mathbf{V}$ corresponding to nonzero eigenvalues are elements of $\Si^{1/2}\Sb$.  Therefore, re-standardizing with respect to $\Si^{-1/2}$ provides a basis for $\Sb$, or at least part of $\Sb$ if the rank is less than $K$.

Similar to OLS, SIR can perform well for various model types, but it also fails when the link function is symmetric around the mean of $\bX$. In this case transformations may correct this issue \citep{Garnham2016} or other dimension reduction methods, such as SAVE \citep{CookWeis91} may be used.  Based on slice variances, in addition to the LDC in Condition \ref{LDC}, the constant conditional variance condition required by SAVE is:
\begin{condition}\label{CCC}
Var$(\mathbf{X}|\mathbf{B}^\top\mathbf{X})$ is constant.
\end{condition}
Conditions \ref{LDC} and \ref{CCC} are satisfied when, for example, $\X$ is normally distributed. 

\section{Slice weighted average regression}\label{sec:SWAR}

The proposed method, which we call Slice Weighted Average Regression (SWAR), was conceptualized by noting the advantages of the slicing approach implemented by SIR and obtaining the slice slope vectors given by OLS. Since the population slope vector of OLS (i.e. the slope to be estimated) is given as $\bols = [\text{Var}(\X)]^{-1} \text{Cov}(\X, Y)$, then this means that SWAR is based on the slice covariances between $Y$ and $\X$, as opposed to the slice means of SIR and slice variances of SAVE.

The obtained slice slope vectors are then weighted and combined into a matrix whose eigenvectors corresponding to non-zero eigenvalues are elements of the CDRS. Within the $h$th slice, the slice slope vector is denoted by, $\bh = \left[\text{Var}(\mathbf{X}|Y\in S_h)\right]^{-1} \text{Cov}\left(\mathbf{X}, Y|Y\in S_h \right)$. As indicated in the following lemma, when both Condition \ref{LDC} and \ref{CCC} hold, this slice slope vector contains information regarding the CDRS.  The proof can be found in Appendix \ref{ap:bhinSb}.

\begin{lem}\label{lemma:bh in Sb}
If Conditions \ref{LDC} and \ref{CCC} hold, then under the model in \eqref{eq:DRmodel},
\begin{equation*}
    \left[\text{Var}(\mathbf{X}|Y\in S)\right]^{-1} \text{Cov}\left(\mathbf{X}, Y|Y\in S\right) \in \Sb 
\end{equation*}
for any subrange $S$ of $Y$.
\end{lem}
 
The SWAR matrix is then given by, 
\begin{equation}\label{eq:V(w)}
\mathbf{R} = \sum^H_{h=1}w_h \bh \bh^\top,
\end{equation}
where $w_h$, $h = 1, \dots, H$ are weights for the slices. E.g, like for SIR we could use $w_h=p_h=P(Y\in S_h)$, however we also consider different weighting choices later.

\begin{thm}\label{thm:EigenVecsinSb}
If Conditions \ref{LDC} and \ref{CCC} hold, then under the model in \eqref{eq:DRmodel},
eigenvectors corresponding to non-zero eigenvalues of $\mathbf{R}$ are elements of $\Sb$.
\end{thm}
\begin{proof}
This follows directly from Lemma \ref{lemma:bh in Sb} since each of the $\bh$'s are elements of the CDRS.
\end{proof}

From Theorem \ref{thm:EigenVecsinSb}, SWAR can return an orthonormal basis for $\Sb$, since the returned eigenvectors do not require a re-standardization, as long as rank$(\mathbf{R}) =K$.  Like OLS and SIR, for some model types SWAR may only find a partial basis, and the associated discussions with OLS and SIR also hold here. 

\begin{remark}
Recall that SIR can find at most $H-1$ e.d.r. directions. With respect to SWAR, max\{rank($\mathbf{R}$)\} = $H$, so that a complete basis may be found when $H \geq K$.  In fact, the special case of $H=1$ is simply the usual OLS slope vector which can be used to determine the direction in the single-index model.  However, as we will see later, combining multiple slopes can be beneficial.
\end{remark}

An advantage of this method, compared to OLS, is the ability to find more than one informative \edr direction for models with $K>1$. As shown in the following sections, SWAR can provide improved estimates compared to other methods in some contexts. Additionally, SWAR can perform well in the presence of contamination, for example, when model or distributional assumptions are violated by some observations in the data. Identifying contaminant points and returning a good \edr direction estimate at the same time is a great advantage of SWAR. The alternative weighting approaches given in later sections can improve the estimation further, providing more robust estimates. These claims are supported by simulation results and an example. 

Consider a sample data set $\{y_i,\mathbf{x}_i\}^n_{i=1}$ and call $(y_i,\mathbf{x}_i)$ the $i$th pair.  Let $n_h$ denote the number of observations to be in the $h$th slice $(h=1,\ldots,H)$ where $n_1+\ldots+n_H=n$.  The SWAR estimating algorithm is defined as follows: 
\begin{description}
    \item[Step 1.] Order the pairs according to the order of the $y_i$s so that the $i$th ordered pair has the $i$th smallest $y_i$. 
    \item[Step 2.] Partition the ordered data into $H$ slices, with $n_h$ observations in the $h$th slice $(h=1,\ldots,H)$.
    \item[Step 3.] Obtain the OLS slope vector estimates for each of the slices and denote these as $\widehat{\bm{b}}_1, \dots, \widehat{\bm{b}}_H$. 
    \item[Step 4.] Form the SWAR matrix,  $\widehat{\mathbf{R}} = \displaystyle\sum^H_{h=1} {w}_h \widehat{\bm{b}}_h \widehat{\bm{b}}_h^\top$, where $w_h = n_h/n$.
    \item[Step 5.] Return the eigenvectors $\widehat{\g}_1, \dots, \widehat{\g}_K$ as the estimated basis for $\Sb$.
\end{description}

A simple slicing strategy that is commonly used for SIR, and can also be used for SWAR, is to choose the number of slices $H$, and then to allocate an equal number (or approximately equal) of observations per slice (i.e., ${w}_h=1/H$).

Throughout this paper, OLS is used to obtain the slice coefficient vectors, however, other \edr direction estimators could easily be utilized as well. For example, \cite{li1989} showed that robust linear regression estimators such as M-estimators can also identify \edr directions. 

\begin{remark}
It is important to note that if OLS is used as the slope estimator, then SWAR cannot be performed if the number of observations in a slice is equal or less than the dimensionality of $\X$, i.e. if $n_h \leq p$.  SIR is not limited by such a case since the slice mean can be determined for any slice with at least one observation.  For SWAR, care needs to be taken that not too many slices are chosen. 
\end{remark}

Some other methods that also combine multiple coefficient vectors into a dimension reduction matrix are the Principal Quantile Regression \citep[PQR,][]{PQR} and the Principal Asymmetric Least Squares \citep[PALS,][]{PALS}, which instead of slices use varying quantile and expectile levels respectively. The main disadvantage of PQR compared to PALS is that it is a computationally intense procedure, with PALS being at least twice as fast as PQR as noted by \cite{PALS}.

Robustness studies of SIR and related methods have shown that outliers can be harmful to estimation \citep[e.g.][]{Gather2002}.  However, not all outliers are influential as shown by example by \cite{SH&MC01}.  As a tool for studying the robustness properties of estimators, influence functions for SIR have shown that it is the direction of the predictor vector relative to the e.d.r. directions that largely determines whether an outlier is influential \citep{PRENDERGAST_2005,PR2007}.  Also, the response only contributes to slice placement for SIR, so an additional consideration for SWAR is to the extent that outlying response values, even in just one slice, can influence estimation overall.  This leads us to the study of influences functions for SWAR for two reasons.  Firstly to better understand the robustness properties of SWAR, and secondly to introduce influence-derived weights for the robustness of SWAR. 

\section{Influence functions for SWAR}\label{sec:IFs}

The Influence Function \citep[IF,][]{Hampel74} measures the relative influence of a contaminant on an estimator of interest. In other words, it measures how much the estimator has changed by the addition or removal of a small amount  of contamination. Consider the following contamination distribution that allows for contamination in both the response and predictor variables, defined as
\begin{equation}
    G_\e = (1-\e)G + \e \Delta_{\bm{w}_0}
\end{equation}
where $0 < \e < 1$ is the proportion of contamination, $G$ is the uncontaminated joint distribution of $(Y, \X)$ and $\Delta_{\bm{w}_0}$ is the Dirac measure, putting all of its mass at the contaminant point $\bm{w}_0 = (y_0, \bm{x}_0)$. For a statistical estimator with functional $T$ defined at $G$ and $G_\e$, the IF in the direction of $\bm{w}_0$ is defined as, 
\begin{equation}
    \text{IF}(T, \bm{w}_0; G) = \lim_{\e \downarrow 0 } \dfrac{T(G_\e) - T(G)}{\e} = \dfrac{\partial T(G_\e)}{\partial \e} \Bigg|_{\e = 0}.
\end{equation}
A contaminant $\bm{w}_0$ is highly influential when there is a big difference between the estimator at $G$ and $G_\e$ resulting in large IF. More information about the influence function can be found, e.g., in \cite{Hampel86} and \cite{clarke2018robustness}.

The following assumption was used in the derivation of the SIR influence functions \citep[e.g.][]{PRENDERGAST_2005} .  This is a realistic assumption that assumes the slicing proportions are the same with and without contamination (e.g. $H$ equally proportioned slices in both cases). 

\begin{assumption}\label{as:indep}
The slicing proportions $w_1,\dots, w_h$ are pre-determined independently of $G$.
\end{assumption}

\subsection{Influence function and asymptotic variance for a single e.d.r. direction}

Let $I(y_0 \in S_h)$ denote the indicator function which is equal to 1 when $y_0$ belongs in the $h$th slice and 0 otherwise.  Also recall that the OLS slope vector for slice $h$ is $\bh$ $(h=1,\ldots,H)$ and $\bm{\Sigma}_h=\text{Var}_G(\X|Y\in S_h)$  is the covariance of the predictor vector in the $h$th slice.  In the case of $K>1$, closed form solutions of the IF for e.d.r directions do not exist.  Therefore, we consider the single-index case here, and in the next section a different IF approach for when $K>1$.

Let $\gamma_1$ denote the functional for the first e.d.r direction estimator where $\gamma_1(G)=\g_1$.  For the single-index model case where $K=1$, the influence function is given below. 

\begin{thm}\label{thm:IFg1}
Let $h_0\in (1,\ldots,H)$ denote the slice within which the contamination is positioned (i.e. $y_0\in S_{h_0}$) and use other notations defined previously. Under the model in \eqref{eq:SIM} with $K=1$ (single-index model) and Assumption \ref{as:indep}, if Condition \ref{LDC} and \ref{CCC} hold, then the influence function for the SWAR e.d.r direction, with functional $\gamma_1$, at $G$ is given by,
\begin{align}
    \text{IF}(\gamma_1, \bm{w}_0; G) &= \dfrac{1}{\lambda_1} \displaystyle\sum_{h=1}^H I(y_0 \in S_h) r_{0,h} (\bm{I}_p -\g_1 \g_1^\top ) \ShInv (\bm{x}_0 - \mh)\bh^\top \g_1  \nonumber\\
    &= \dfrac{r_{0,h_0}}{\lambda_1}  (\bm{I}_p - \bm{P} ) \Si^{-1} (\bm{x}_0 - \bm{\mu})\bm{b}_{h_0}^\top \g_1
\end{align}
where $r_{0,h_0}=y_0 - E_G(Y|Y\in S_{h_0}) - \bh^\top(\bm{x}_0-\bm{\mu}_h)$ is the OLS residual for the contaminant within the $h_0$th slice and $\mathbf{P}=\g_1\g_1^\top$ is the projection matrix onto the CDRS. 
\end{thm}

The proof of Theorem \ref{thm:IFg1} is given in Appendix \ref{ap:IFVw}.  Note that $\mathbf{I}-\mathbf{P}$ is a projection matrix onto the compliment of the CDRS.  This highlights that it is the direction of the predictor vector that can largely determine influence and, as with SIR and other methods, explains why not all outliers are influential.  We will look at some specific examples later.  

For an estimator with functional $T$ that is sufficiently regular, so that $T(G_n)$ is asymptotically normal, then the asymptotic variance of the estimator, ASV$(T,G)$, at $G$ is equal to \citep[see, e.g.][]{Hampel86},
\begin{equation}\label{eq:ASVformula}
    \text{ASV}(T, G) = \text{E}_G [\text{IF}(T, \bm{W};G) \text{IF}(T, \bm{W};G)^\top] 
\end{equation}
where $\bm{W}$ is a random variable.

For a normal $\X$, the asymptotic variance of the SIR \edr direction estimator, denoted by $b_{SIR}$, is given by \cite{PRENDERGAST_2005} as, ASV$(b_{SIR}, G) = \big(\frac{3}{2}- \frac{1}{\nu_1}\big)\bm{\eta}_1 \bm{\eta}_1^\top + \big(\frac{1}{\nu_1} - 1\big) \Si^{-1}$ . Here, $\bm{\eta}_1$ is the SIR re-standardized eigenvector that corresponds to the largest non-zero eigenvalue of the SIR matrix, denoted as $\nu_1$. 

The following theorem gives the ASV of the SWAR \edr direction, the proof of which is given in Appendix \ref{ap:ASVproof}.

\begin{thm}\label{thm:ASVswar}
For a random $\X$, using previously introduced notation and when Condition \ref{LDC} and \ref{CCC} hold, the asymptotic variance of the SWAR \edr direction estimate under the model in \eqref{eq:SIM} is given by,
\begin{equation}
    ASV(\gamma_1, G) = \dfrac{1}{\lambda_1^2} \displaystyle\sum_{h=1}^H w_h (\bh^\top \g_1)^2 \text{E}(R_h^2|Y \in S_h) (\bm{I} - \bm{P})\Si^{-1} (\bm{I} - \bm{P})
    \end{equation}
    where $R_h=Y - E(Y|Y\in S_h)-\bh^\top(\X - \bm{\mu}_h)$.
\end{thm}

The ASV($\gamma_1, G$) is a $p\times p$ symmetric matrix whose diagonal elements are the ASVs of the $p$ elements of the SWAR \edr direction estimate and the off-diagonal elements are the asymptotic covariances between the elements. 

\subsection{Influence function for the subspace estimator}

In dimension reduction we are mainly interested in the \edr direction estimators. However, \cite{PRENDERGAST_2005} showed that an observation may be influential on a particular \edr direction but have no influence on the corresponding \edr space. Therefore, an influence function for the dimension reduction space estimator is more appropriate than it is for individual e.d.r. directions.
Since SWAR returns an orthonormal basis, a candidate measure of influence is introduced by \cite{Benasseni1990SensitivityCF} in the context of principal components.  This measure is the average length of the distance vector between each principal component and its projection onto the space spanned by the contaminated components.  In the context of SWAR, let $\bm{\Gamma}$ denote the $p\times K$ matrix whose columns $\g_1, \dots, \g_K$ are the \edr directions given at $G$. Similarly, $\bm{\Gamma}_\e$ is the corresponding matrix at $G_\e$. Then the measure of distance between the contaminated and the uncontaminated \edr spaces is
\begin{equation}
    r(\bm{\Gamma}, \bm{\Gamma}_\e) = 1 - \dfrac{1}{K} \displaystyle\sum_{k = 1}^K ||(\bm{I}_p - \bm{P}(\e)) \g_k||
\end{equation}
where $\bm{P}(\e) = \bm{\Gamma}_\e \bm{\Gamma}_\e^\top$ is the projection matrix onto the subspace spanned by the contaminated \edr directions $\g_1(\e), \dots, \g_K(\e)$. By letting $\rho$ denote the functional for B\'{e}nass\'{e}ni's measure, the influence function of $\rho$ for $\bm{w}_0$ at $G$, is given by,
\begin{equation}
    \text{IF}(\rho, \bm{w}_0 ; G) = \lim_{\e \downarrow 0} \dfrac{ r(\bm{\Gamma}, \bm{\Gamma}_\e)  - 1}{\e}
\end{equation}

Given that the $\g_k$'s and the $\g_k(\e)$'s, for  $k=1,\dots,K$ are orthogonal and have unit length, then it is clear that there is no influence on the \edr space estimator when $r(\bm{\Gamma}, \bm{\Gamma}_{\e}) = 1$, which happens when span$(\g_1,\dots, \g_K) = \text{span}(\g_1(\e), \dots, \g_K(\e))$. Then, the influence is at its highest at $r(\bm{\Gamma}, \bm{\Gamma}_{\e}) = 0$ which occurs when the aforementioned spans are orthogonal to each other. 

\begin{thm}\label{thm:IFrho}
Under Assumption \ref{as:indep} and given that Condition \ref{LDC} and \ref{CCC} hold, the influence function for B\'{e}nass\'{e}ni's measure applied to the SWAR \edr space at $G$ is given by,
\begin{equation}
    \text{IF}(\rho, \bm{w}_0; G) =  -\dfrac{|r_{0,h_0}|}{K}\left(\sum^K_{k=1}\dfrac{|\g_k^\top \bm{b}_{h_0}|}{\lambda_k}\right)\left\|(\bm{I}_p-\bm{P})\Si^{-1}(\bm{x}_0-\bm{\mu})\right\|
\end{equation}
\end{thm}

The proof for Theorem \ref{thm:IFrho} is given in Appendix \ref{ap:BenasProof}. From Theorem \ref{thm:IFg1} and \ref{thm:IFrho}, it is clearly evident that for $K = 1$, $\text{IF}(\rho, \bm{w}_0; G) = - \| \text{IF}(\gamma_1, \bm{w}_0; G)\|$. 

\subsection{Some examples}

\subsubsection{Identifying influence of certain observation types}

Through the influence functions given in Theorem \ref{thm:IFg1} and \ref{thm:IFrho}, we can identify how certain types of observations affect the estimation of the \edr direction and the \edr space provided by SWAR. For example, we observe the below interesting cases. Note that both IFs are functions of the residuals, $r_{0,h_0}$, and $(\bm{I}_p-\bm{P})\Si^{-1}(\bm{x}_0-\bm{\mu})$, so that the cases that follow are true for both the \edr direction and the \edr space estimators. 

\textit{Case 1: Contaminant equal to the mean, $\bm{x}_0 = \bm{\mu}$} \\
When $\bm{x}_0 = \bm{\mu}$, there is zero influence on the aforementioned estimators. It is interesting, that in this case there is no effect from $y_0$ on the influence, even if $y_0$ has an extreme value or violates the model. 

\textit{Case 2: $\Si^{-1} (\bm{x}_0 - \bm{\mu})$ is an element of the \edr space}\\
If $\Si^{-1} (\bm{x}_0 - \bm{\mu}) \in \Sb$, we again have zero influence on the estimators since $(\bm{I}_p-\bm{P})$ is the projection matrix onto the complement of the \edr space and so, $(\bm{I}_p-\bm{P}) \Si^{-1} (\bm{x}_0 - \bm{\mu}) = 0$.

\textit{Case 3: $\Si^{-1} (\bm{x}_0 - \bm{\mu})$ is orthogonal to the \edr space}\\
For this observational type the IFs increase without bound as the length of $\Si^{-1} (\bm{x}_0 - \bm{\mu})$ increases. For example, let $\Si = \bm{I}_p$ and $\bm{\mu}=0$, then we have $\text{IF}(\gamma_1, \bm{w}_0; G) = \frac{y_0 - \mu_{y,h_0}}{\lambda_1} (\bm{I}_p - \bm{P}) \bm{x}_0 \bbeta_{h_0}^\top \g_1$ and $\text{IF}(\rho, \bm{w}_0; G) = -\frac{|y_0 - \mu_{y,h_0}|}{K} \left(\sum^K_{k=1}\frac{|\g_k^\top \bm{b}_{h_0}|}{\lambda_k}\right) \left\|(\bm{I}_p-\bm{P}) \bm{x}_0 \right\|$. Therefore, an observation $(y_0, \bm{x}_0)$ can have unbounded influence on the estimators if $y_0$ and/or $\bm{x}_0$ are arbitrarily large. 

\textit{Case 4: The residuals of $\bm{w}_0$ are equal to zero.}\\
There is also zero influence from the contaminant $\bm{w}_0$ on either of the estimators if $r_{0,h_0} = 0$.

\subsubsection{Influence function plots}\label{subsec:IFplots}

\begin{figure}[h]
   \hfill
   \subfigure[IF$(\gamma_{1,1}, \bm{w}_0; G)$]{\includegraphics[width=7.5cm]{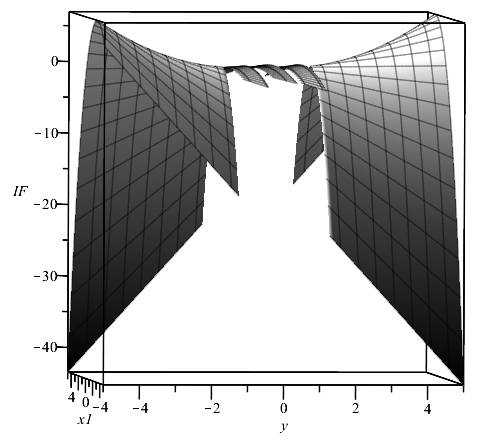}}
   \hfill
   \subfigure[IF$(\rho, \bm{w}_0; G)$]{\includegraphics[width=8cm]{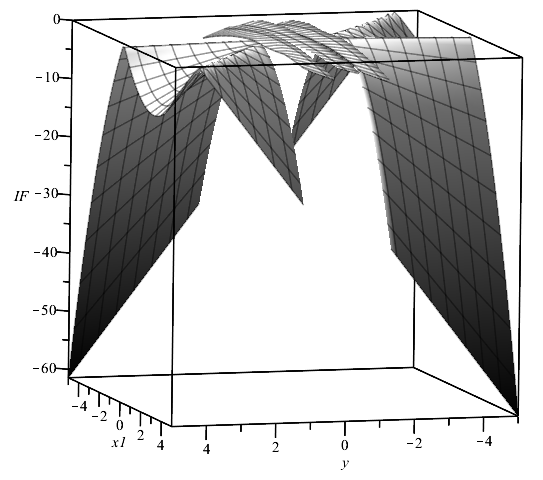}}
   \hfill
   \caption{Plots of IF$(\gamma_{1,1}, \bm{w}_0; G)$ and IF$(\rho, \bm{w}_0; G)$ for the model in \eqref{mod:IFplotsExample} with $\bm{x}_0 = [x_1, 0]$ and varying $y_0$ and $x_1$.}
   \label{fig:IFplots1}
\end{figure}

Below we provide some example influence plots to visually demonstrate the effect of a contaminant on the \edr direction and the \edr space estimators. Consider the linear model
\begin{equation}\label{mod:IFplotsExample}
    Y = \bbeta^\top\X + 0.5\e,
\end{equation}
with $\bbeta= [1, -1]^\top$, $\X \sim N_2(\bm{0}, \bm{I}_2)$, $\e \sim N(0,1)$, and $H=5$ equally probable slices. Let IF$(\gamma_{1,1}, \bm{w}_0; G)$ denote the influence value of the contaminant $\bm{w}_0$ on the first element of the \edr direction $\g_1$ given by SWAR. In Figure \ref{fig:IFplots1}, plot (a) is the $\text{IF}(\g_{1,1}, \bm{w}_0; G)$ and plot (b) is $\text{IF}(\rho, \bm{w}_0; G)$, where for $\bm{x}_0 = [x_1, x_2]$, we set $x_2 = 0$ and allow for $y_0$ and $x_1$ to vary. In both plots there is zero influence on the estimators when $x_1$ is also zero, which is expected as explained in \textit{Case 1}. Within each slice, as $x_1$ and $y_0$  increase, the influence on $\gamma_{1,1}$ in plot (a) increases without bound along the diagonal, that is when $y_0$ moves towards the boundaries that determine the slice sub-ranges.  In plot (b) of Figure \ref{fig:IFplots1}, the influence on the \edr space follows the same trends as in plot (a) but we also observe approximately zero influence values when $y_0$ follows the model approximately, i.e. when $y_0 \approx x_1$ since $x_2 = 0$.

\begin{figure}
    \hfill
    \subfigure[IF$(\gamma_{1,1}, \bm{w}_0; G)$ View 1]{\includegraphics[width = 7.5cm]{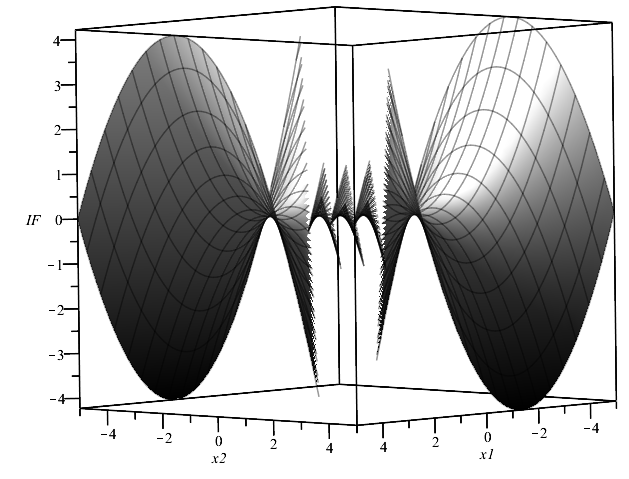}}
    \hfill
    \subfigure[IF$(\gamma_{1,1}, \bm{w}_0; G)$ View 2]{\includegraphics[width = 7.5cm]{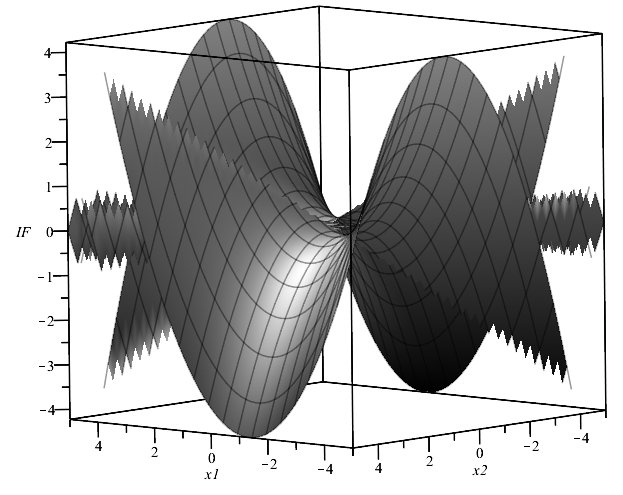}}
    \hfill
    \caption{Plots of IF$(\gamma_{1,1}, \bm{w}_0; G)$ shown from two different angles, for the model in \eqref{mod:IFplotsExample}, with $y_0 = \bbeta^\top \bm{x}_0$ and varying $x_1$ and $x_2$.}
    \label{fig:IFplots2}
    \end{figure}

\begin{figure}
    \hfill
    \subfigure[IF$(\rho, \bm{w}_0; G)$ View 1]{\includegraphics[
    width = 7.7cm]{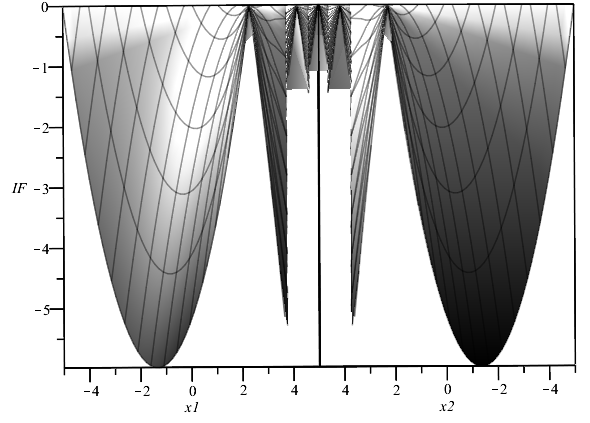}}
    \hfill
    \centering
    \subfigure[IF$(\rho, \bm{w}_0; G)$ View 2]{\includegraphics[width = 7.5cm]{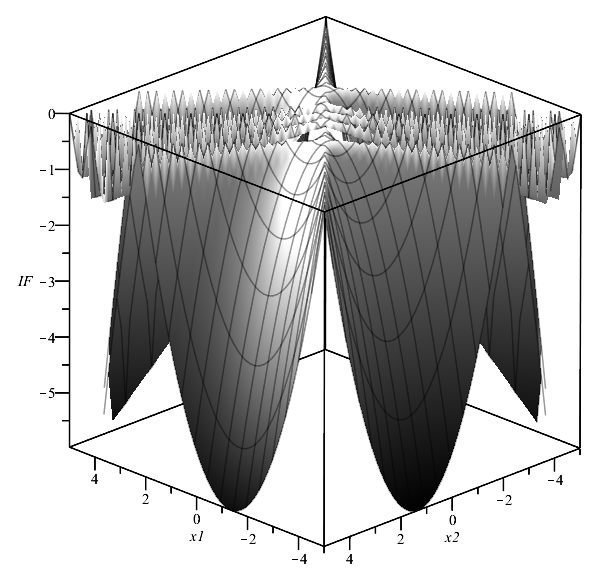}}
    \hfill
    \caption{Plots of the IF$(\rho, \bm{w}_0; G)$, shown from two different angles, for the model in \eqref{mod:IFplotsExample} with $y_0 = \bbeta^\top \bm{x}_0$, $\bm{x}_0 = [x_1, x_2]$ and for varying $x_1$ and $x_2$. }
    \label{fig:IFplots3}
\end{figure}

In Figure \ref{fig:IFplots2} and \ref{fig:IFplots3} we provide two different views for each of the $\text{IF}(\gamma_{1,1}, \bm{w}_0; G)$ and $\text{IF}(\rho, \bm{w}_0; G)$, respectively, where for the model in \eqref{mod:IFplotsExample} we let the response contaminant be consistent with the model, i.e. $y_0 = \bbeta^\top \bm{x}_0$, and vary $x_1$ and $x_2$. In both Figures the influence on the estimators is zero when $\bm{x}_0 = [0,0]^\top = \bm{\mu}$, as expected. Now, for the model in \eqref{mod:IFplotsExample}, $\bm{x}_0$ is an element of $\Sb$ for any $c\in \mathbb{R}$ where $\bm{x}_0 = c\bbeta = [c, -c]^\top$. Therefore, when $x_2 = -x_1$ then we expect to have zero influence on both estimators, as explained in \textit{Case 2} earlier. This is clearly shown in plots (b) of Figure \ref{fig:IFplots2} and \ref{fig:IFplots3}. Furthermore, due to the nature of the model, $y_0 = x_1 - x_2$, when both $x_1$ and $x_2$ increase simultaneously, for example, from 0 to 5, or when their difference is fairly small, then we fall in the middle slices where the influence is relatively small. Although influence still increases within those slices as we reach their limits. The influence increases (towards the far positive or negative values) without bound when the absolute value of the difference ($x_1 - x_2$) is large.  


\subsection{Sample and Empirical influence functions}\label{subsec:SIF_EIF}

In the sample setting, the influence of the $i$th observation is found by considering the change of the estimator pre- and post-removal of the observation. For a sample of size $n$ denoted by $\{y_i, \bm{x}_i\}_{i=1}^n$, denote the empirical distribution by $G_n$ and the empirical distribution without the $i$th observation by $G_{n,(i)}$. Then, the sample influence function (SIF) of an estimator with functional $T$ is given by $\text{SIF}(T, \bm{w}_i ; G_n) = (n-1) \big[ T(G_n) - T(G_{n,(i)})\big]$.  Therefore, the SIF for the SWAR \edr direction estimator, with functional $\gamma_k$, at $G_n$ is given by, 
\begin{equation}
    \text{SIF}(\gamma_k, \bm{w}_i; G_n) = (n-1)[\widehat{\g}_k - \widehat{\g}_{k,(i)}]
\end{equation}
for $k = 1,\dots, K$. Hence, large values in the components of the $\text{SIF}(\gamma_k, \bm{w}_i; G_n)$ vector indicate that observation $i$ is highly influential.

Similarly, for B\'{e}nass\'{e}ni's measure, let $\widehat{\bm{\Gamma}}$ and $\widehat{\bm{\Gamma}}_{(i)}$ denote the matrices whose columns are the estimated \edr directions at $G_n$ and $G_{n,(i)}$. Then, the sample influence function for $\rho$ for the $i$th observation at $G_n$ is
\begin{equation}\label{eq:SIFrho}
    \text{SIF}(\rho, \bm{w}_i; G_n) = (n-1)\big[r(\widehat{\bm{\Gamma}}, \widehat{\bm{\Gamma}}_{(i)}) - 1\big].
\end{equation}
It is clear that when $|\text{SIF}(\rho, \bm{w}_i; G_n)|$ is large, then the $i$th observation is highly influential on the basis that is estimating the CDRS. 

The disadvantage of the SIF is that it requires $(n+1)$ estimates in order to obtain the sample influence values for the entire sample which can be computationally expensive especially when $n$ and/or $p$ are large.  To overcome this, the empirical influence function (EIF) is an approximation of the SIF and  can be found by replacing the population parameters of the IF with their sample estimates. Hence,  the EIF of the $i$th observation for $\rho$ for SWAR is 
\begin{equation}
    \text{EIF}(\rho, \bm{w}_i; G_n) = - \dfrac{|\widehat{r}_{i,h}|}{K} \Bigg( \sum^K_{k=1}\dfrac{|\widehat{\g}_k^\top  \widehat{\bm{b}}_{h}|}{\widehat{\lambda}_k}\Bigg)\left\|(\mathbf{I}_p-\widehat{\mathbf{P}}) \widehat{\Si}^{-1} (\bm{x}_i-\widehat{\bm{\mu}})\right\|,
\end{equation}
where $\bm{w}_i$ belongs in the $h$th slice. For large $n$, EIF $\approx$ SIF and it can be used in practise to efficiently detect influential observations.  Below we provide a comparison between the SIF and EIF of B\'{e}nass\'{e}ni's measure for SWAR. 

\begin{figure}[h]
    \centering
    \includegraphics[width=0.95\textwidth, page=1]{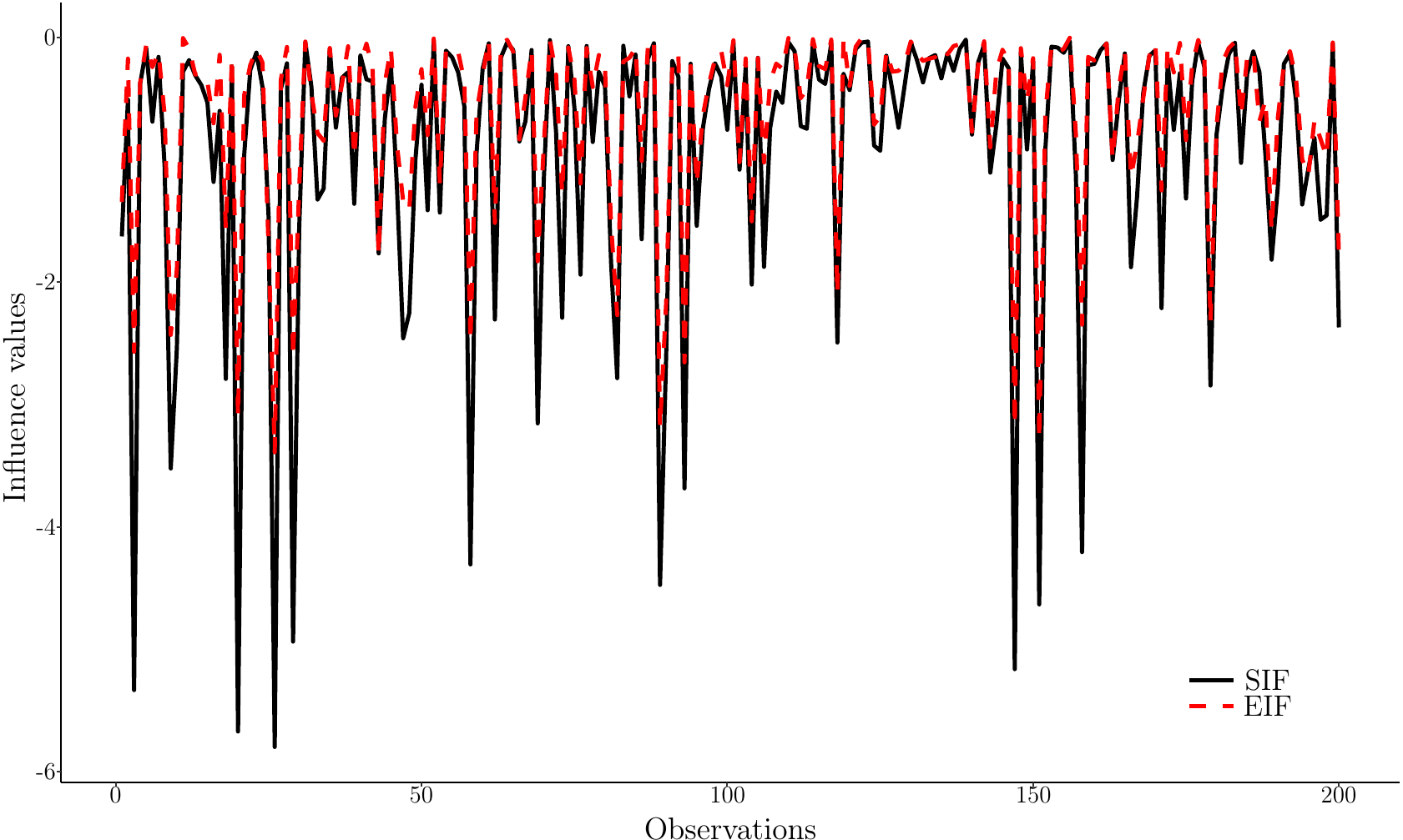}
    \caption{Sample and Empirical influence values on the \edr space estimator of each observation for the model given in \eqref{mod:IFplotsExample}.}
    \label{fig:SIFvsEIF}
\end{figure}

Consider the model as defined in \eqref{mod:IFplotsExample} with $n = 200$ and $p = 5$. The SIF and EIF values for each of the observations are depicted in Figure \ref{fig:SIFvsEIF}. The figure shows that the EIF is a good approximation to the SIF and so can be used to identify the most influential observations without repeated SWAR estimation.

\section{Some applications of the influence function}\label{sec:IFweights}

The usefulness of the IF expands further than just being a tool to explore the robustness properties of an estimator. In this section, we propose using the mean influence to choose optimal weights for the slices in SWAR, as well as choosing $K$ and $H$. 

Our first goal is to down-weigh the slices with high mean influence allowing the slices with more stable estimates to contribute more to SWAR. We consider two different weighting techniques: namely, the within slice mean influence and the total mean influence. Detailed explanations for each of these are given below. 

\subsection{Within slice mean influence weights}\label{within}

Let $\X_n$ denote the $n\times p$ matrix whose $i$th row is $\bm{x}_i^\top$. Also, let $\widehat{\bm{b}}_{h,(i)}$ denote the $h$th slice slope vector without the $i$th observation and $\X_{n_h}$ be the matrix whose rows are the $\bm{x}_j^\top$'s that fall in the $h$th slice.  For this version of weighting, we consider influence on the dimension reduced predictors within the $h$th slice, and where the dimension reduction has been carried out using the estimated slope vectors within that slice.  Similar approach to determining influential observations have been considered previously \citep[e.g.][]{Prendergast2008, Smith10}.   For each of the observations within the $h$th slice, compute
\begin{equation}\label{eq:SIFols}
    \delta_{h, i} = (n_h-1)^2 \Big[1- \text{cor}^2( \X_{n_h}\widehat{\bm{b}}_{h}, \X_{n_h,(i)}\widehat{\bm{b}}_{h,(i)}) \Big]\;\;\;(i=1,\ldots,n_h) .
\end{equation}
Now, let $\bar{\delta_h}=n_h^{-1}\sum^{n_h}_{i=1}\delta_{h,i}$ be the mean of the $\delta_{h, i}$s.  Then, the new weights are defined as follows, 
\begin{equation}\label{eq:weightW}
    w_h=\displaystyle\frac{1}{\bar{\delta_h}}\cdot \frac{1}{\|\widehat{\bm{b}}_h\|^2}\;\;\; (h=1,\ldots,H).
\end{equation}
and where these are scaled to sum to one.

From here onwards, this version of SWAR which uses the within slice mean influence weights, given in \eqref{eq:weightW}, will be referred to as SWAR$_W$.

\subsection{Total mean influence weights}\label{total}

For this is a re-weighting process first calculate the SWAR estimate (see Section \ref{sec:SWAR}) and then calculate the SIF in \eqref{eq:SIFrho} for each of the $n$ observations. Now, let $\{\rho_{h,i}\}^{n_h}_{i=1}$ denote the SIFs for the observations in the $h$th slice and $|\overline{\rho}_h|=|n_h^{-1}\sum^{n_h}_{i=1}\rho_{h,i}|$ the absolute sample mean of these values. Then, the total mean influence weights are given by, 
\begin{equation}\label{eq:weightT}
    w_h = \dfrac{1}{|\overline{\rho}_h|}\cdot \dfrac{1}{\|\bm{b}_h \|^2}.
\end{equation}

We then recompute the SWAR estimate but where these weights are used instead.  The new re-weighted SWAR method with the total mean influence weights will be referred to as SWAR$_T$. 

It is important to note that for SWAR$_W$ and SWAR$_T$, the term $1/\|\bm{b}_h\|^2$ is used in the weights so that the $h$th slice slope vectors are normalised. This is because  only the direction of the vectors is important here and a vector with large length can dominate the dimension reduction matrix and affect the \edr direction estimates. Therefore, transforming the slope vectors to have unit length solves this problem.

\subsection{Choosing $K$ and $H$ in SWAR}\label{sec:IFforK}

Influence functions can have multiple applications in the dimension reduction setting. Another application is that of choosing dimension reduction parameters (i.e. $H$ and $K$) based on minimum mean influence, for example. Such applications have been explored in \cite{shakerthesis}, and a similar approach has been presented by \cite{YeWeiss} and \cite{LiquetSaracco, LiquetSaracco12}, who evaluated the sensitivity of dimension reduction estimators at different parameter values using a bootstrap approach.

In the mean influence approach, which we adopt here, we consider the optimal pair $H$, $K$ to be the one that results in the minimum mean influence. For SWAR, the sample influence function of B\'{e}nass\'{e}ni's measure can be used to select $H$ in conjunction with $K$.  An example of this, for a simulated model, is given soon in Section \ref{sec:simulations}. An inspection of the Estimated SSPs (ESSPs) is also recommended when choosing $H$ and $K$ in practice.  

The only disadvantage of the mean SIF compared to the bootstrap approaches is the computational intensity of the SIF. However, for large sample sizes where EIF $\approx$ SIF, the EIF can be used in practice to choose $H$ and $K$ more efficiently, since computational intensity reduces significantly. The HIF \citep{PR2007} provides another alternative to the SIF which can provide a better approximation than the EIF and is also less time consuming than the SIF. The effectiveness of the SIF in choosing $K$ and $H$ is not examined thoroughly in the present paper. 

\section{Simulations}\label{sec:simulations}

In this section, we provide simulated examples to demonstrate the effectiveness of SWAR, SWAR$_W$ and SWAR$_T$. The performance of the aforementioned methods is compared with those of OLS, PALS  and SIR and results provided.  We have chosen principal asymmetric least squares \citep[PALS,][]{PALS} since it is an interesting new method that also combines least squares estimates.  The PALS object function is $$L_\tau(\beta_0,\bbeta_1)=\bbeta_1^\top \Si \bbeta_1 + \lambda \text{E}[\rho_{\tau}(R^*)]$$
where $R^*=Y - \beta_0-\bbeta_1^\top(\bX-\bm{\mu})$ and 
where $\rho_\tau$ is the asymmetric least squares loss function \citep{newey1987asymmetric}.  For given values for $\tau\in [0,1]$, a slope using the above objective function is computed (estimated on the standardised scale and pre-multiplied by $\widehat{\Si}^{-1/2}$), and combined similarly to our Step 4 of SWAR but with weights all equal to one.

For every model, we perform 1000 repetitions for each combination of the sample sizes $n = 50, 200, 500$ and $1000$ with $p = 5, 10$ and $20$, and $H = 2, 5$ and $10$ for the slicing methods. For PALS we follow the lead of \cite{PALS} and set the tuning parameter $\lambda = 1$ and $\tau = 0, 0.1, 0.2, \dots,0.9, 1$. To measure and evaluate the performance of each method we use the squared canonical correlations between the true and estimated \edr spaces.

\subsection{Single-Index Model} 

Consider the following model with $\X \sim N_p(\bm{0}, \bm{I}_p)$ independent of $\e$ and $\e \sim N(0,1)$, 

\begin{model}\label{mod:ContNoCont}
\quad $Y = \bX + (1 + 0.7\bX + 0.6\e)^3$,  
\end{model}
where the true $\bbeta = [-1 , 2, 0, -1, 0, \dots, 0]$. 

\begin{figure}[h]
    \centering
    \includegraphics[width = 0.85\textwidth, page =2]{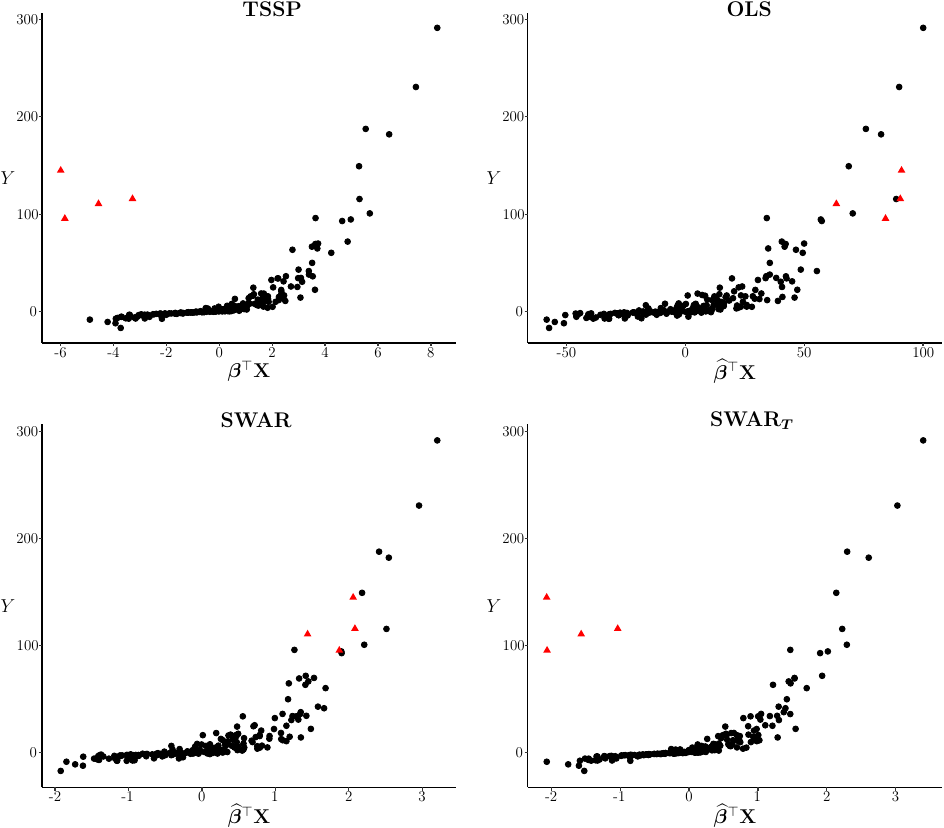}
    \caption{The True (top left) and Estimated SSPs for a realization of the contaminated Model \ref{mod:ContNoCont} with $n = 200$ and $p = 10$. The ESSPs of OLS (to right), SWAR (bottom left) and SWAR$_T$ (bottom right) are depicted in the figure.}
    \label{fig:ESSPsCont}
\end{figure}

For this model we first perform the aforementioned methods for each combination of $n, p$ and $H$. Then, we introduce contamination and perform the same comparisons to see where the influence weighting strategies can improve estimation. We then replace $2\%$ of the ordered observations based on the value of $Y$ with contamination, where the contamination response is generated from $N(150, 30^2)$ and we subtract five from each of the associated predictor vector elements. 

To highlight the effect of this type of contamination, we first provide some example ESSPs in Figure \ref{fig:ESSPsCont} for $n=200$ and $p=10$.   The contamination is clearly evident in the true ESSP (i.e. assuming known $\bbeta$).  However, from the estimated ESSPs using OLS and SWAR, we can no longer distinguish these points from other observations.  However, the estimated ESSP using SWAR$_T$ provides an excellent estimate of the ESSP.  Hence we have chosen this type of contamination since resulting ESSPs may provide no indication that some problematic observations need further inspection. 

\begin{table}[h]
    \centering
     \caption{Average values of the squared correlations, $\text{cor}(\bbeta^\top \X_n, \widehat{\bbeta}^\top \X_n)^2$, with the corresponding standard deviations in parentheses, for 1000 realizations of Model \ref{mod:ContNoCont} given by OLS, PALS, SIR, SWAR, SWAR$_W$ and SWAR$_T$, for the uncontaminated and contaminated case.}
    \label{tab:Mod1_tables}
    \textbf{Uncontaminated}
    \resizebox{\textwidth}{!}{\begin{tabular}{l@{\hskip 16pt}lll@{\hskip 16pt}lll@{\hskip 16pt}lll}
    \toprule
        \multirow{2}{*}{Method} & \multicolumn{3}{c}{$n = 50$} & \multicolumn{3}{c}{$n = 200$} & \multicolumn{3}{c}{$n = 500$}  \\ 
        ~ & $p = 5$ & $p = 10$ & $p = 20$ & $p = 5$ & $p = 10$ & $p = 20$ & $p = 5$ & $p = 10$ & $p = 20$ \\ \toprule
        OLS & 0.953 (0.034) & 0.900 (0.051) & 0.811 (0.062) & 0.986 (0.010) & 0.969 (0.016) & 0.936 (0.023) & 0.994 (0.004) & 0.986 (0.007) & 0.972 (0.010) \\ [6pt]
        PALS ($\lambda = 1$) & 0.947 (0.039) & 0.890 (0.057) & 0.802 (0.066) & 0.983 (0.012) & 0.962 (0.020) & 0.925 (0.029) & 0.993 (0.005) & 0.983 (0.009) & 0.965 (0.014) \\ [6pt]
        SIR ($H = 2$) & 0.949 (0.033) & 0.890 (0.046) & 0.796 (0.055) & 0.987 (0.009) & 0.971 (0.014) & 0.941 (0.017) & 0.995 (0.003) & 0.988 (0.005) & 0.976 (0.008) \\ 
        SIR ($H = 5$) & 0.984 (0.012) & 0.962 (0.020) & 0.921 (0.031) & 0.996 (0.003) & 0.992 (0.004) & 0.983 (0.006) & 0.999 (0.001) & 0.997 (0.001) & 0.993 (0.002) \\ 
        SIR ($H = 10$) & 0.988 (0.009) & 0.972 (0.016) & 0.937 (0.031) & 0.998 (0.002) & 0.995 (0.003) & 0.989 (0.004) & 0.999 (0.001) & 0.998 (0.001) & 0.996 (0.001) \\[6pt] 
        SWAR ($H = 2$) & 0.967 (0.029) & 0.914 (0.063) & 0.641 (0.215) & 0.992 (0.006) & 0.982 (0.010) & 0.960 (0.017) & 0.997 (0.002) & 0.993 (0.004) & 0.984 (0.006) \\ 
        SWAR ($H = 5$) & 0.863 (0.168) & $-$ & $-$ & 0.983 (0.014) & 0.957 (0.028) & 0.875 (0.069) & 0.994 (0.005) & 0.985 (0.008) & 0.966 (0.014) \\ 
        SWAR ($H = 10$) & $-$ & $-$ & $-$ & 0.956 (0.053) & 0.848 (0.132) & $-$ & 0.986 (0.011) & 0.966 (0.021) & 0.909 (0.046) \\ [6pt]
        SWAR$_W$ ($H = 2$) & 0.980 (0.017) & 0.942 (0.039) & 0.709 (0.192) & 0.995 (0.004) & 0.988 (0.006) & 0.972 (0.011) & 0.998 (0.002) & 0.995 (0.003) & 0.989 (0.004) \\ 
        SWAR$_W$ ($H = 5$) & 0.960 (0.075) & $-$ & $-$ & 0.998 (0.002) & 0.994 (0.005) & 0.982 (0.014) & 0.999 (0.001) & 0.998 (0.001) & 0.996 (0.002) \\
        SWAR$_W$ ($H = 10$) & $-$ & $-$ & $-$ & 0.993 (0.007) & 0.978 (0.036) & $-$ & 0.998 (0.002) & 0.995 (0.003) & 0.987 (0.007) \\[6pt] 
        SWAR$_T$ ($H = 2$) & 0.966 (0.033) & 0.910 (0.071) & 0.656 (0.214) & 0.988 (0.009) & 0.975 (0.015) & 0.944 (0.026) & 0.995 (0.004) & 0.988 (0.006) & 0.975 (0.011) \\
        SWAR$_T$ ($H = 5$) & 0.943 (0.147) & $-$ & $-$ & 0.996 (0.017) & 0.987 (0.051) & 0.968 (0.087) & 0.999 (0.001) & 0.998 (0.003) & 0.996 (0.006) \\
        SWAR$_T$ ($H = 10$) & $-$ & $-$ & $-$ & 0.936 (0.147) & 0.901 (0.169) & $-$ & 0.975 (0.084) & 0.965 (0.088) & 0.937 (0.121) \\ \bottomrule
    \end{tabular}}%
    \vspace{10pt}
    \textbf{Contaminated}
    \resizebox{\textwidth}{!}{\begin{tabular}{l@{\hskip 16pt}lll@{\hskip 16pt}lll@{\hskip 16pt}lll}
    \toprule
    \multirow{2}{*}{Method} & \multicolumn{3}{c}{$n = 50$} & \multicolumn{3}{c}{$n = 200$} & \multicolumn{3}{c}{$n = 500$} \\ 
        ~ & $p = 5$ & $p = 10$ & $p = 20$ & $p = 5$ & $p = 10$ & $p = 20$ & $p = 5$ & $p = 10$ & $p = 20$ \\ \toprule
        OLS & 0.308 (0.190) & 0.289 (0.170) & 0.238 (0.145) & 0.337 (0.110) & 0.301 (0.108) & 0.271 (0.097) & 0.340 (0.077) & 0.297 (0.073) & 0.275 (0.067) \\ [6pt]
        PALS ($\lambda = 1$) & 0.321 (0.193) & 0.301 (0.171) & 0.244 (0.146) & 0.357 (0.113) & 0.320 (0.111) & 0.285 (0.099) & 0.362 (0.080) & 0.320 (0.077) & 0.296 (0.070) \\ [6pt]
        SIR ($H=2$) & 0.785 (0.071) & 0.720 (0.068) & 0.627 (0.072) & 0.812 (0.033) & 0.773 (0.033) & 0.732 (0.034) & 0.816 (0.021) & 0.784 (0.021) & 0.758 (0.022) \\ 
        SIR ($H=5$) & 0.773 (0.062) & 0.737 (0.061) & 0.687 (0.068) & 0.788 (0.029) & 0.754 (0.030) & 0.726 (0.031) & 0.789 (0.019) & 0.755 (0.019) & 0.733 (0.020) \\ 
        SIR ($H=10$) & 0.741 (0.069) & 0.707 (0.068) & 0.661 (0.074) & 0.762 (0.031) & 0.723 (0.033) & 0.694 (0.034) & 0.766 (0.020) & 0.723 (0.021) & 0.698 (0.022)\\ [6pt]
        SWAR ($H=2$) & 0.382 (0.203) & 0.404 (0.182) & 0.285 (0.170) & 0.441 (0.112) & 0.461 (0.104) & 0.463 (0.092) & 0.449 (0.077) & 0.467 (0.070) & 0.480 (0.062)  \\ 
        SWAR ($H=5$) & 0.401 (0.233) & $-$ & $-$ & 0.454 (0.117) & 0.471 (0.112) & 0.429 (0.117) & 0.470 (0.074) & 0.504 (0.068) & 0.518 (0.060) \\
        SWAR ($H=10$) & $-$ & $-$ & $-$ & 0.473 (0.181) & 0.379 (0.177) & $-$ & 0.502 (0.100) & 0.491 (0.094) & 0.451 (0.100)  \\ [6pt]
        SWAR$_W$ ($H=2$) & 0.951 (0.064) & 0.879 (0.122) & 0.562 (0.257) & 0.948 (0.050) & 0.878 (0.093) & 0.825 (0.102) & 0.960 (0.029) & 0.900 (0.052) & 0.854 (0.068)  \\ 
        SWAR$_W$ ($H=5$) & 0.916 (0.147) & $-$ & $-$ & 0.984 (0.035) & 0.970 (0.037) & 0.944 (0.059) & 0.992 (0.008) & 0.983 (0.016) & 0.973 (0.021) \\ 
        SWAR$_W$ ($H=10$) & $-$ & $-$ & $-$ & 0.976 (0.058) & 0.955 (0.055) & $-$ & 0.991 (0.011) & 0.982 (0.020) & 0.969 (0.028)  \\[6pt] 
        SWAR$_T$ ($H=2$) & 0.954 (0.057) & 0.895 (0.103) & 0.605 (0.253) & 0.990 (0.009) & 0.977 (0.019) & 0.954 (0.026) & 0.996 (0.004) & 0.991 (0.006) & 0.982 (0.011) \\ 
        SWAR$_T$ ($H=5$) & 0.900 (0.198) & $-$ & $-$ & 0.988 (0.055) & 0.976 (0.071) & 0.954 (0.091) & 0.997 (0.007) & 0.996 (0.004) & 0.992 (0.010) \\ 
        SWAR$_T$ ($H=10$) & $-$ & $-$ & $-$ & 0.880 (0.196) & 0.853 (0.207) & $-$ & 0.932 (0.153) & 0.924 (0.145) & 0.884 (0.188) \\ \bottomrule
    \end{tabular}}%
\end{table}

\begin{figure}[h]
    \centering
    \includegraphics[width = \textwidth, page=3]{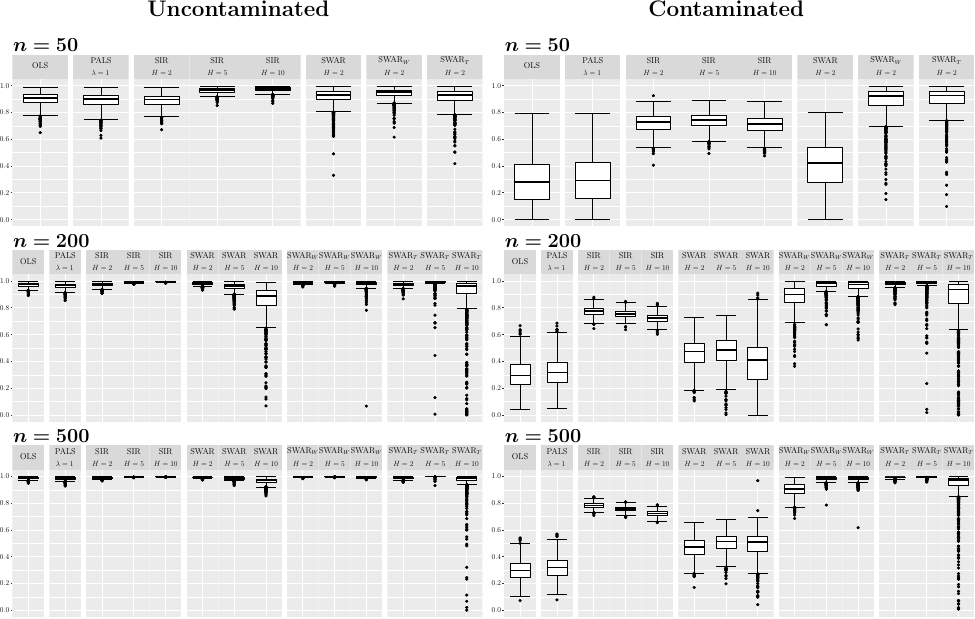}
    \caption{Box-plots of the squared correlations between the true and estimated \edr spaces given by (1000 repetitions of) OLS, PALS (with $\lambda =1$), SIR, SWAR, SWAR$_W$ and SWAR$_T$, for the uncontaminated (left) and contaminated (right) Model \ref{mod:ContNoCont}, with $n = 50, 200$ and $500$, $H = 2, 5$ and $10$ and $p = 10$. }
    \label{fig:BoxsCont_noCont}
\end{figure}

The averages of the squared correlations along with their corresponding standard deviations, shown in parentheses, for the uncontaminated and contaminated Model \ref{mod:ContNoCont}, are given in Table \ref{tab:Mod1_tables}. The missing values (denoted by $-$) for SWAR, SWAR$_W$ and SWAR$_T$ in the tables are for when $p \geq n_h$ so that slopes cannot be estimated. Figure \ref{fig:BoxsCont_noCont}, shows the box-plots of the correlations from 1000 repetitions of each method for $p = 10$. For simplicity, the results for $n = 1000$ are omitted from the table and the box-plots since they display similar trends with $n = 500$.

We can see that all methods perform extremely well for the uncontaminated case, especially as the sample size increases. OLS and PALS tend to perform better than SWAR methods when the sample size is small and dimensionality is large (e.g., $n = 50$ and $p = 20$). SIR is also better than SWAR for small sample sizes but performance is similar for larger $n$. Whereas SIR's performance is better when the number of slices is large, all versions of SWAR prefer a smaller or intermediate number of slices.

For the contaminated Model \ref{mod:ContNoCont}, the performance of OLS and PALS deteriorates significantly. SWAR also performs poorly, but to a lesser degree, and SIR provides reasonable estimates but still a notable decrease in performance. However, the weighted adjusted methods SWAR$_W$ and SWAR$_T$ show an outstanding performance in the presence of contamination, which was shown by example in Figure \ref{fig:ESSPsCont} for SWAR$_T$.  

\begin{table}[h]
\centering
\caption{Choices based on the SIF of the SWAR \edr space estimator for number of slices $H$ and dimension reduced predictors $K$, for the uncontaminated Model \ref{mod:ContNoCont}.}
\label{tab:SIFchoices}
\resizebox{\textwidth}{!}{\begin{tabular}{ll@{\hskip 18pt}rrr@{\hskip 18pt}rrr@{\hskip 18pt}rrr@{\hskip 18pt}rrr}
    \toprule
        ~ & ~ & \multicolumn{3}{c}{$n = 50$} & \multicolumn{3}{c}{$n = 200$} & \multicolumn{3}{c}{$n = 500$} & \multicolumn{3}{c}{$n = 1000$} \\
        ~ & ~ & $p = 5$ & $p = 10$ & $p = 20$ & $p = 5$ & $p = 10$ & $p = 20$ & $p = 5$ & $p = 10$ & $p = 20$ & $p = 5$ & $p = 10$ & $p = 20$ \\ \toprule
        \multirow{3}{*}{$K = 1$} & $H = 2$ & 918 & 1000 & 988 & 760 & 905 & 995 & 637 & 799 & 941 & 515 & 686 & 862 \\ 
         & $H = 5$ & 82 & $-$ & $-$ & 186 & 83 & 5 & 298 & 183 & 57 & 403 & 274 & 133 \\ 
         & $H = 10$ & $-$ & $-$ & $-$ & 54 & 12 & $-$ & 65 & 18 & 2 & 82 & 40 & 5 \\ [9pt]
        \multirow{3}{*}{$K = 2$} & $H = 2$ & 0 & 0 & 12 & 0 & 0 & 0 & 0 & 0 & 0 & 0 & 0 & 0 \\ 
         & $H = 5$ & 0 & $-$ & $-$ & 0 & 0 & 0 & 0 & 0 & 0 & 0 & 0 & 0 \\ 
         & $H = 10$ & $-$ & $-$ & $-$ & 0 & 0 & $-$ & 0 & 0 & 0 & 0 & 0 & 0 \\ \bottomrule
    \end{tabular}}
\end{table}

Finally, Table \ref{tab:SIFchoices} shows how many times (out of 1000 repetitions) the SIF given in \eqref{eq:SIFrho} has chosen each pair of $H$ and $K$ for SWAR, for the uncontaminated Model \ref{mod:ContNoCont}. Between the choice of $K=1$ and $K=2$, the SIF is correctly choosing $K=1$ in all replications except for $n = 50$ and $p=20$ for which case the method is struggling to find a good estimate. For the optimal number of slices $H = 2$ is chosen the most. This is consistent with the previous findings that SWAR prefers a smaller number of slices.  As $n$ increases, a larger number of slices is often chosen.

\subsection{Multiple-Index model}\label{subsec:MIMexample}

Similarly to the previous example, for $\X \sim N_p(\bm{0}, \bm{I})$ independent of $\e$ with $\e \sim N(0,1)$, consider now 
\begin{model}\label{mod:MIMexample}
\quad $Y = 2 + \bbeta_1^\top \X + (1 + 0.5 \bbeta_2^\top \X)^3 + 0.3\e$,  
\end{model}
with true \edr directions, $\bbeta_1 = [1,2,-3, 0, \dots, 0]$ and $\bbeta_2 = [1,1,0,-2, 0, \dots, 0]$. 

\begin{table}[!ht]
    \centering
        \caption{Average values of the squared canonical correlations between the true and estimated directions, with the corresponding standard deviations in parentheses, from 1000 repetitions given by PALS (with $\lambda=1$ and $\lambda=200$), SIR, SWAR, SWAR$_W$ and SWAR$_T$ for Model \ref{mod:MIMexample}.}
    \label{tab:MIMeg}
    \resizebox{\textwidth}{!}{\begin{tabular}{l@{\hskip 18pt}lll@{\hskip 18pt}lll@{\hskip 18pt}lll}
    \toprule
    \multirow{2}{*}{Method} & \multicolumn{3}{c}{$n = 50$} & \multicolumn{3}{c}{$n = 200$} & \multicolumn{3}{c}{$n = 500$}  \\ 
        ~ & $p = 5$ & $p = 10$ & $p = 20$ & $p = 5$ & $p = 10$ & $p = 20$ & $p = 5$ & $p = 10$ & $p = 20$ \\
        \toprule
         PALS ($\lambda = 1$) & 0.668 (0.381) & 0.577 (0.417) & 0.508 (0.421) & 0.722 (0.341) & 0.617 (0.403) & 0.560 (0.435) & 0.786 (0.288) & 0.680 (0.354) & 0.598 (0.410)  \\ 
        PALS ($\lambda = 200$) & 0.772 (0.301) & 0.667 (0.354) & 0.583 (0.393) & 0.915 (0.131) & 0.829 (0.202) & 0.729 (0.282) & 0.960 (0.061) & 0.906 (0.114) & 0.828 (0.188) \\[6pt]
        SIR ($H = 2$) & 0.510 (0.471) & 0.489 (0.436) & 0.455 (0.384) & 0.512 (0.485) & 0.513 (0.472) & 0.514 (0.446) & 0.513 (0.487) & 0.517 (0.481) & 0.528 (0.463) \\ 
        SIR ($H = 5$) & 0.763 (0.312) & 0.638 (0.370) & 0.545 (0.398) & 0.932 (0.108) & 0.852 (0.178) & 0.752 (0.261) & 0.976 (0.036) & 0.940 (0.073) & 0.880 (0.127)  \\
        SIR ($H = 10$) & 0.746 (0.329) & 0.617 (0.393) & 0.529 (0.422) & 0.930 (0.123) & 0.843 (0.200) & 0.736 (0.290) & 0.978 (0.033) & 0.945 (0.066) & 0.888 (0.121) \\ [6pt]
        SWAR ($H = 2$) & 0.905 (0.168) & 0.807 (0.242) & 0.589 (0.353) & 0.972 (0.054) & 0.935 (0.090) & 0.867 (0.157) & 0.988 (0.022) & 0.970 (0.043) & 0.936 (0.078) \\ 
        SWAR ($H = 5$) & 0.924 (0.158) & $-$ & $-$ & 0.987 (0.032) & 0.961 (0.072) & 0.884 (0.170) & 0.994 (0.011) & 0.984 (0.027) & 0.961 (0.059) \\ 
        SWAR ($H = 10$) & $-$ & $-$ & $-$ & 0.977 (0.060) & 0.904 (0.177) & $-$ & 0.988 (0.026) & 0.966 (0.062) & 0.900 (0.154)  \\ [6pt]
        SWAR$_W$ ($H = 2$) & 0.905 (0.168) & 0.807 (0.242) & 0.589 (0.353) & 0.972 (0.054) & 0.935 (0.090) & 0.867 (0.157) & 0.988 (0.022) & 0.970 (0.043) & 0.936 (0.078) \\ 
        SWAR$_W$ ($H = 5$) & 0.934 (0.154) & $-$ & $-$ & 0.993 (0.011) & 0.979 (0.028) & 0.930 (0.098) & 0.997 (0.005) & 0.992 (0.010) & 0.980 (0.022)  \\ 
        SWAR$_W$ ($H = 10$) & $-$ & $-$ & $-$ & 0.990 (0.034) & 0.954 (0.102) & $-$ & 0.997 (0.005) & 0.992 (0.010) & 0.976 (0.032) \\[6pt]
        SWAR$_T$ ($H = 2$) & 0.905 (0.168) & 0.807 (0.242) & 0.589 (0.353) & 0.972 (0.054) & 0.935 (0.090) & 0.867 (0.157) & 0.988 (0.022) & 0.970 (0.043) & 0.936 (0.078) \\ 
        SWAR$_T$ ($H = 5$) & 0.795 (0.305) & $-$ & $-$ & 0.955 (0.103) & 0.869 (0.208) & 0.687 (0.350) & 0.988 (0.021) & 0.962 (0.067) & 0.902 (0.138) \\ 
        SWAR$_T$ ($H = 10$) & $-$ & $-$ & $-$ & 0.743 (0.348) & 0.584 (0.410) & $-$ & 0.871 (0.244) & 0.726 (0.347) & 0.578 (0.420)  \\ \bottomrule
    \end{tabular}}
\end{table}

\begin{figure}[h]
    \centering
    \includegraphics[width = \textwidth, page = 4]{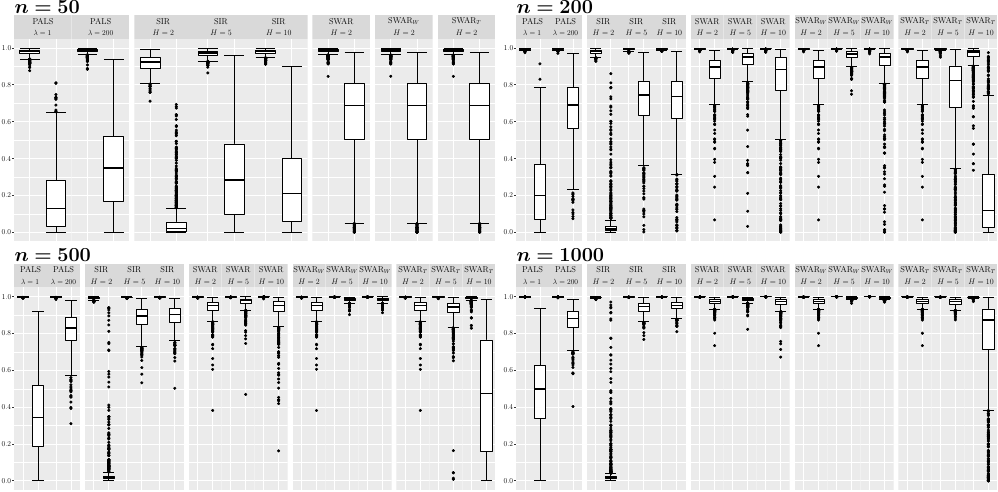}
    \caption{Box-plots of the squared canonical correlations for $1000$ repetitions from PALS with $\lambda=1$ and $\lambda = 200$, SIR, SWAR, SWAR$_W$ and SWAR$_T$, for the various $n$ and $H$ combinations with $p = 10$, for Model \ref{mod:MIMexample}.}
    \label{fig:MIMboxplots}
\end{figure}

The average values of the squared canonical correlations with the corresponding standard deviations, in parentheses, can be found in Table \ref{tab:MIMeg}. The box-plots of the squared canonical correlations for $p = 10$ are then given in Figure \ref{fig:MIMboxplots}.  In the boxplots, we provide the squared canonical correlations for each direction so that we can compare performance for both directions.

For this model, PALS with $\lambda = 1$ fails to find a second direction, however, a different value of the tuning parameter seems to improve the results significantly especially as $n$ increases.  Therefore, this model is an example case where $\lambda$ can have a significant effect on estimation, a perhaps unexpected result based on analysis carried out about PALS so far, where $\lambda$ did not have a significant effect on estimation \citep{PALS}. SIR struggles to find a good estimate of the subspace when $n$ is small (for $H=2$ this is expected since the SIR matrix is of rank 1), and it can often find it difficult to estimate the second direction unless $n$ is large, as is evident in the box-plots. 

All SWAR versions perform comparatively well even when the sample size is small, in which case there is more variability in the estimation of the second direction. The performance of the SWAR methods also declines slightly when $p$ increases. Additionally, SWAR and SWAR$_T$ benefit from a small to intermediate choice of $H$, whereas SWAR$_W$ benefits from an intermediate to a larger number of slices.

\section{\textit{BigMac} data example}\label{sec:example}

In this section we compare the estimated directions given by OLS, SIR, SWAR, SWAR$_W$ and SWAR$_T$ for the \textit{`BigMac'} data from \cite{Enz}. Of interest is the regression of the response, the minimum labor required to buy a Big Mac and fries from MacDonalds in each city, on the 9 socio-economic predictor variables. There are 45 observations in the data and so we choose $H=5$ for SIR and for SWAR, SWAR$_W$ and SWAR$_T$ we choose $H = 2$. These choices of $H$ was confirmed to be the most suited to each method through the inspection of the Estimated SSPs (ESSPs). Inspections of ESSPs also revealed that $K=1$ was suitable. 

\begin{figure}[h!]
    \centering
    \includegraphics[width = 0.84\textwidth, page = 5]{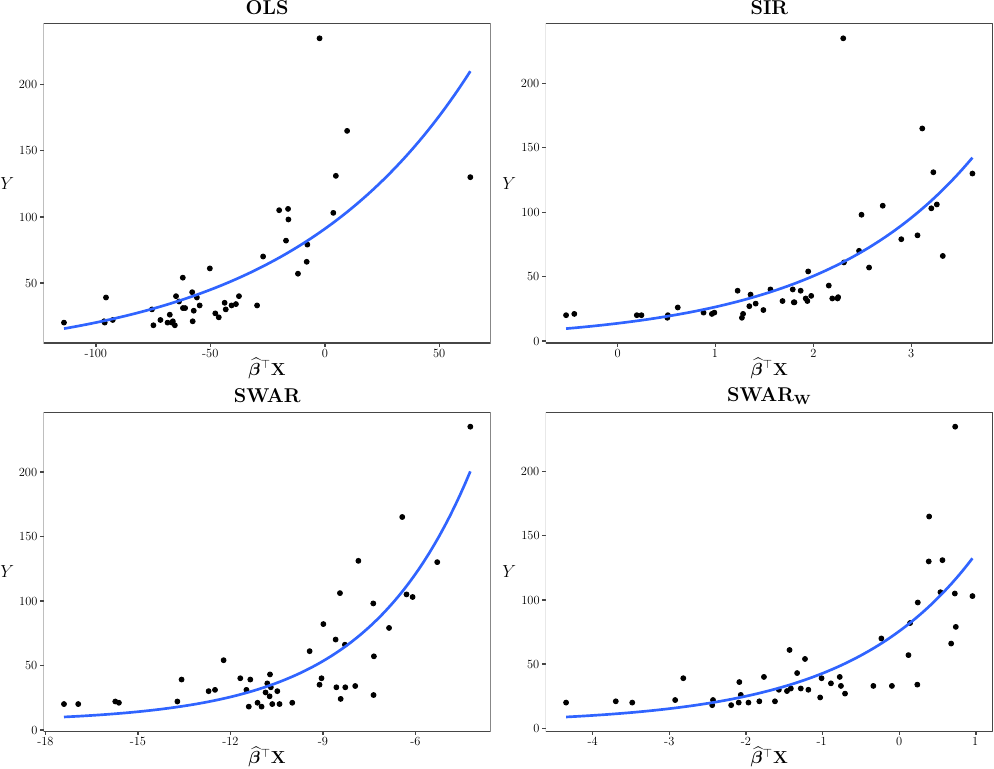}
    \caption{ESSPs given by OLS, SIR, SWAR and SWAR$_W$ for the \textit{`BigMac'} data set.}
    \label{fig:BigmacEssp}
\end{figure}

The ESSPs for each method are shown in Figure \ref{fig:BigmacEssp} where an exponential curve has been added to each plot. It is evident from Figure \ref{fig:BigmacEssp} that SWAR provides the best ESSP for the BigMac data with the curve fitting all observations reasonably well. OLS performs comparatively poorly, SIR performs well except for a notable outlier that does not fit well, and SWAR$_W$ also provides a good fit. 

\section{Discussion} \label{sec:discussion}

In this paper we consider slice weight average regresion (SWAR), along with robustified versions of this method using influence re-weighting (SWAR$_W$ and SWAR$_T$).  These versions utilize the mean sample influence function to down-weight the slices that contain highly influential observations which therefore provide a non-robust estimate. This weighting process can also be easily applied to other dimension reduction methods that combine multiple vectors into a dimension reduction matrix. Although not reported here, we also considered re-weighted versions of SIR and PALS but with no notable improvements (some slight improvements for PALS was detected). 

Despite the fact the OLS can only be used for a single direction $(K=1)$, SWAR method is surprisingly capable of finding a second direction for the models we considered.

\noindent \textbf{Acknowledgments.} We are very grateful to Abdul-Nasah Soale and Yuexiao Dong for kindly providing the implementation code for PALS. 

\section*{Declarations}

\noindent \textbf{Funding.} This research is funded by the Australian Government’s Research Training
Program (RTP) in support of MM’s doctoral research degree at La Trobe University. The
funding body had no role in the reported research and in writing the manuscript.

\noindent \textbf{Code availability.} R codes are available from the corresponding author upon request.

\noindent \textbf{Data availability.} The public data set, used in this article, and additional information can be found at \url{https://www.rdocumentation.org/packages/ldr/versions/1.3.3/topics/bigmac} .

\noindent \textbf{Conflicts of interest.} The authors have no conflicts of interest to declare that are relevant to the content of this article.

\begin{appendices}
\section{Proof of Lemma \ref{lemma:bh in Sb}}\label{ap:bhinSb}

Let $\mathbf{B}=[\bbeta_1,\ldots,\bbeta_K]$, $S$ be any subrange of the range of $Y$ and, for simplicity in what follows,  $\bm{C}_S=\text{Var}(\bm{X}|Y\in S)-\Si$. First, from Lemma 1 of \cite{PR2007}, when Conditions \ref{LDC} and \ref{CCC} hold,
\begin{equation}\label{eq:Varx}
    \text{Var}(\mathbf{X}|Y\in S) = \Si + \Si \B \left(\B^\top \Si \B \right)^{-1} \B^\top \bm{C}_S \B \left(\B^\top \Si \B \right)^{-1} \B^\top \Si.
\end{equation}
 Also, from \cite{PRENDERGAST_2005}, under Condition \ref{LDC}
\begin{equation}\label{result2}
E(\mathbf{X}|\mathbf{B}^\top\mathbf{X})=\bm{\mu}+\bm{\Sigma}\mathbf{B}\left(\mathbf{B}^\top\bm{\Sigma}\mathbf{B}\right)^{-1}\mathbf{B}^\top(\mathbf{X}-\bm{\mu}).
\end{equation}

Then, it is simple to show, from the Woodbury matrix identity \citep{woodbury-1950}, that,
\begin{equation}\label{inverse_varS}
\left[\text{Var}(\mathbf{X}|Y\in S)\right]^{-1}=\Si^{-1} - \B \left(\B^\top \Si \B \right)^{-1} \left[(\B^\top \mathbf{C}_S \B)^{-1} + \left(\B^\top \Si \B \right)^{-1} \right]^{-1} \left(\B^\top \Si \B \right)^{-1}\B^\top.
\end{equation}

Furthermore, by conditioning and \eqref{result2}, we get,
\begin{align}\label{eq:CovXYinS}
\text{Cov}\left(\mathbf{X}, Y|Y\in S\right)=& \E\left\{\left[Y - E(Y|Y\in S)\right]\left[\mathbf{X}-E(\mathbf{X}|Y\in S)\right]|Y\in S\right\}\nonumber\\
=&\E \left[ \left\{ Y - E(Y|Y \in S)\right\} \left\{ \E(\X|\BX) - \E[\E(\X|\BX)|Y\in S)]\right\} |Y \in S\right] \nonumber\\ 
=& \text{Cov}(\m + \C (\X - \m), Y |Y \in S) \nonumber\\
=& \C \text{Cov}(\X, Y|Y \in S)
\end{align}
for any range $S$ of $Y$, since $\E(\X|Y\in S) = \E[\E(\X|\BX)|Y\in S]$ due to the independence between $\e$ and $\X$. 

Finally, from \eqref{inverse_varS} and \eqref{eq:CovXYinS}, $\left[\text{Var}(\mathbf{X}|Y\in S)\right]^{-1}\text{Cov}(\mathbf{X}, Y|Y\in S)$ can be written in the form 
$$ \B (\B^\top \Si \B)^{-1} \left[ \bm{I} - \left( \left( \B^\top \bm{C}_S \B \right)^{-1} \B^\top \Si \B + \bm{I}_K \right)^{-1} \right] \B^\top \text{Cov}(\X, Y|Y\in S).$$  This is then an element of $\Sb$ which completes the proof.

\section{Influence functions derivations for SWAR} 

\subsection{Influence functions for the SWAR matrix and the single-index model e.d.r. direction}\label{ap:IFVw}

Let $C_h$ and $C_{XY,h}$ denote the functionals for the $h$th slice covariance matrices so that at $G$, $C_h(G) = \bm{\Sigma}_h$ and $C_{XY,h}(G) = \bm{\Sigma}_{xy,h}$ respectively.  In addition, let $B_h$ denote the functional of the $h$th slice slope vector estimator where $B_h(G) = C_h^{-1}(G) C_{XY,h}(G) = \ShInv \Si_{xy,h}$.  For simplicity, we let $\Tilde{\bm{x}}_0 = \bm{x}_0 - \mh$ and $\Tilde{y}_0 = y_0 - \mu_{y,h}$.  From \cite{PR2007}, 
\begin{equation}\label{eq:IFCh}
    \text{IF}(C_h, \bm{w}_0; G) = \dfrac{\partial }{\partial \e} \text{var}_G\big[x|y \in S_h(\e)\big] \bigg|_{\e=0} - \dfrac{1}{w_h} I(y_0 \in S_h) \big[  \bm{\Sigma}_h - \Tilde{\bm{x}}_0 \Tilde{\bm{x}}_0^\top \big].
\end{equation}

Since $C_h^{-1}(\Ge) C_h(\Ge) = \bm{I}_p$, using the Product Rule and by setting $\e$ to 0 we have,
$\text{IF}( C_h^{-1}, \bm{w}_0; G) = - \ShInv \text{IF}( C_h, \bm{w}_0; G) \bm{ \Sigma}_h^{-1}$.  Then, from \eqref{eq:IFCh}, the influence function for the inverse slice variance is   
\begin{equation}\label{eq:IFCh-1}
    \text{IF}(C_h^{-1}, \bm{w}_0; G) = \dfrac{1}{w_h} I(y_0 \in S_h) \ShInv \Big[ \Si_h - \Tilde{\bm{x}}_0 \Tilde{\bm{x}}_0^\top \Big]\ShInv - \ShInv \dfrac{\partial }{\partial \e}\text{var}_G[x|y \in S_h(\e)] \bigg|_{\e=0} \ShInv.
\end{equation}

For brevity here we omit the details, but by following closely the proof for \eqref{eq:IFCh} \citep[see Lemma A1 proof of][]{PR2007}, we have
\begin{equation}\label{IFsCxy}
    \text{IF}(C_{XY,h}, \bm{w}_0; G) =  \dfrac{\partial}{\partial \e} \text{cov}_G[\bm{x},y|y \in S_h(\e)] \bigg|_{\e=0} + \dfrac{1}{w_h} I(y_0 \in S_h) \Big[ \Tilde{y}_0 \Tilde{\bm{x}}_0 - \Sxyh \Big].
\end{equation}

Using the Product Rule and setting $\e$ to 0 to obtain the IF of $B_h$, gives
\begin{align}
    \text{IF}(B_h, \mathbf{w}_0; G) =& \dfrac{\partial C_h^{-1}(G_\e)}{\partial \e}\Bigg|_{\e=0} \Si_{xy,h} + \Si_h^{-1} \dfrac{\partial C_{XY,h}(G_\e)}{\partial \e} \Bigg|_{\e=0}\nonumber\\
    =&   \dfrac{1}{w_h} I(y_0 \in S_h) \ShInv \Big[ \Si_h - \Tilde{\bm{x}}_0 \Tilde{\bm{x}}_0^\top \Big]\Sxyh - \ShInv \dfrac{\partial }{\partial \e}\text{var}_G[x|y \in S_h(\e)] \bigg|_{\e=0} \Sxyh \nonumber\\
+& \bm{\Sigma}_h^{-1}\left[\dfrac{\partial}{\partial \e} \text{cov}_G[\bm{x},y|y \in S_h(\e)] \bigg|_{\e=0} + \dfrac{1}{w_h} I(y_0 \in S_h) \Big( \Tilde{y}_0 \Tilde{\bm{x}}_0 - \Sxyh \Big)\right]\label{IFBh}
\end{align}
using \eqref{eq:IFCh-1} and \eqref{IFsCxy}.

Let $R$ denote the function for the SWAR matrix estimator such that $R(G)=\sum^H_{h=1}w_hB_h(G)B_h^\top(G)$.  Using the Product Rule, the influence function for $R$ is straightforward when using \eqref{IFBh}.  We present the result in the following lemma for use later.

\begin{lem}\label{lem:IFVw}
Under Assumption \ref{as:indep}, the influence function of the SWAR dimension reduction matrix estimator, with functional $R$, at $G$ is given by, 
\begin{align}\label{eq:IFR}
    \text{IF} (R, \bm{w}_0 ; G) &= \displaystyle\sum_{h=1}^H I(y_0\in S_h)  r_{0,h} \ShInv \Tilde{\bm{x}}_0 \bh^\top + \displaystyle\sum_{h=1}^H I(y_0\in S_h) r_{0,h} \bh  \Tilde{\bm{x}}_0^\top \ShInv \nonumber\\
    &\hspace{1.3cm} + \displaystyle\sum_{h=1}^H w_h \ShInv \dfrac{\partial}{\partial \e} \Big\{ \text{Cov}_G[\X,Y|Y\in S_h(\e)] - \text{Var}_G[\X|Y\in S_h(\e)] \bh \Big\} \Bigg|_{\e = 0} \bh^\top \nonumber\\
    &\hspace{1.3cm} + \displaystyle\sum_{h=1}^H w_h \bh \dfrac{\partial}{\partial \e} \Big\{ \text{Cov}_G[Y, \X |Y\in S_h(\e)] - \bh^\top \text{Var}_G[\X|Y\in S_h(\e)] \Big\} \Bigg|_{\e=0} \ShInv
\end{align}
where $r_{0,h}=y_0 - \mu_{y,h}-\bm{b}_h^\top(\bm{x}_0-\bm{\mu}_h)$ is the OLS residual of $\bm{w}_0$ in the $h$th slice, $\Tilde{\bm{x}}_0 = \bm{x}_0 - \mh$, and $\mu_{y,h}$ and $\mh$ are the means of the $y_i$'s and the $\bm{x}_i$'s in the $h$th slice, respectively.
\end{lem}

Eq. 13 of \cite{PRENDERGAST_2005} provides the influence functions for the eigenvectors of the SIR matrix estimator and this result can be adapted and used directly here.  However, from Lemma \ref{lem:IFVw} and in the case $K>1$, the influence functions for SWAR e.d.r. direction estimators depend on the expressions involving $\text{Cov}_G[\x,y|y\in S_h(\e)]$ and $\text{Var}_G[\x|y\in S_h(\e)]$ which cannot be derived further. We therefore focus our attention on the case of $K=1$.  This is of the form, since $\lambda_j = 0$ for $j>1$,
\begin{equation}\label{eq:IFg1}
    \text{IF}(\gamma_1, \bm{w_0}; G) = \dfrac{1}{\lambda_1}\sum\limits_{j=2}^p \g_j \g_j^\top \text{IF}(R, \bm{w_0}; G) \g_1.
\end{equation}

The form of the IF in the case of $K=1$ simplifies due to several results which we list here: (i) from Lemma \ref{lemma:bh in Sb}, we have that $\bh=c_1\bbeta_1$ for some $c_1\in \mathbb{R}$.  Since $\g_1$ is also a scalar multiple of $\bbeta_1$, then $\g_j^\top\bh=0$ for $j\geq 2$ (since $\g_j^\top\g_1=0$); (ii) from \eqref{inverse_varS}, we know that $\Si_h^{-1}=\Si^{-1}-c_2\bbeta_1\bbeta_1^\top$ for a $c_2\in \mathbb{R}$.  Therefore, $\g_j^\top\Si_h^{-1}=\g_j^\top\Si^{-1}$; (iii) from \eqref{eq:Varx} and \eqref{eq:CovXYinS}, $\text{Var}_G[\X|Y\in S_h(\epsilon)]=\Si + c_3\Si\bbeta_1\bbeta_1^\top\Si$ and $\text{Cov}_G[\X,Y|Y\in S_h(\epsilon)]=c_4\Si\bbeta_1$, $c_3,c_4\in \mathbb{R}$.  Hence, using (i)-(iii), only the first term of IF$(R,\bm{w}_;G)$ in Lemma \ref{lem:IFVw} remains in \eqref{eq:IFg1}.  

The proof of Theorem \ref{thm:IFg1} is complete when applying (ii) to this first term also.

\subsection{Proof of Theorem \ref{thm:IFrho}}\label{ap:BenasProof}

From \cite{Benasseni1990SensitivityCF} the influence function for $\rho$, with respect to SWAR, is of the form, 
\begin{equation}\label{eq:IFRpre}
    \text{IF}(\rho, \bm{w}_0 ; G) = -\dfrac{1}{K} \displaystyle\sum_{k = 1}^K \Big|\Big| \dfrac{1}{\lambda_K} (\bm{I} - \bm{P}) \text{ IF}(R , \bm{w}_0 ; G) \g_k \Big|\Big|
\end{equation}

Note that, $(\bm{I} - \bm{P})$ is a projection matrix on the compliment of the CDRS.  Hence, similar to the proof for Theorem \ref{thm:IFg1} where we noted several simplifications, firstly we have $(\bm{I} - \bm{P})\bh=\mathbf{0}$.  Secondly, since $(\bm{I} - \bm{P})\Si_h^{-1}=(\bm{I} - \bm{P})\Si^{-1}$, then using \eqref{eq:CovXYinS} for the form of $\text{Cov}_G(\X,Y|Y\in S)$ for any range $S$
 \begin{align*}
    (\bm{I} - \bm{P})\Si_h^{-1}\text{Cov}_G[\X,Y|Y\in S(\epsilon)] =&(\bm{I} - \bm{P})\Si^{-1}\text{Cov}_G[\X,Y|Y\in S(\epsilon)]\\
    =& (\bm{I} - \bm{P})\Si^{-1}\Si\bm{B}(\bm{B}^\top\Si\bm{B})^{-1}\bm{B}^\top\text{Cov}_G[\X,Y|Y\in S(\epsilon)]\\
    =&\mathbf{0}.
\end{align*}
It can similarly be shown that 
$$(\bm{I} - \bm{P})\Si_h^{-1}\text{Var}_G[\X|Y\in S(\epsilon)]\bh=\mathbf{0}$$
by using the form of the slice variance matrix in \eqref{eq:Varx}. Then, the proof is complete by substituting \eqref{eq:IFR} in \eqref{eq:IFRpre} and using the above simplifications.

\section{Proof of ASV}\label{ap:ASVproof}

For a random $\bm{W}=(Y,\X)$, from Theorem \ref{thm:IFg1} we have
$$\text{IF}(\gamma_1,\bm{W};G)=\frac{1}{\lambda_1}\sum^H_{h=1}I(Y\in S_h)R_h(\bm{I}_p-\bm{P})\Si_h^{-1}(\X-\mu_h)\bh^\top\g_1$$
where $R_h=Y-\mu_{y,h}-\bh^\top(\X-\mu_h)$.  Hence, since $\text{ASV}(\g_1 ,G) = \text{E} [ \text{ IF}(\gamma_1,\bm{W}; G )  \text{ IF}(\gamma_1,\bm{W}; G )^\top ]$ then
\begin{align}\label{eq:ASVsemi}
    ASV(g_1, G) &= \dfrac{1}{\lambda_1^2} \displaystyle\sum_{h=1}^H(\bh^\top \g_1)^2 (\bm{I} - \bm{P} ) \Si^{-1}  \text{E}\Big[ R_h^2 (\X - \m) (\X - \m)^\top |Y\in S_h\Big] \Si^{-1} (\bm{I} - \bm{P}).
\end{align}

Since $Y$ is a function of $\bbeta_1^\top\X$ and $\epsilon$, and $\epsilon$ is independent of $\X$, then by conditioning we can write $E\Big[ R_h^2 (\X - \m) (\X - \m)^\top |Y\in S_h\Big]=E\Big[ R_h^2 E\left\{(\X - \m) (\X - \m)^\top|\bbeta_1^\top\X\right\} |Y\in S_h\Big].$  From \eqref{result2} and since $(\bm{I}-\bm{P})\bbeta_1=\bm{0}$, we can write
\begin{align}
   (\bm{I}-\bm{P})\Si^{-1}& E\Big[ R_h^2 E\left\{(\X - \m) (\X - \m)^\top|\bbeta_1^\top\X\right\} |Y\in S_h\Big]\Si^{-1}(\bm{I}-\bm{P})\nonumber\\
   =& (\bm{I}-\bm{P})\Si^{-1} E\Big[ R_h^2 E\left\{[\X - E(\X|\bbeta_1^\top\X)] [\X - E(\X|\bbeta_1^\top\X)]^\top|\bbeta_1^\top\X\right\} |Y\in S_h\Big]\Si^{-1}(\bm{I}-\bm{P}).\label{E2}
\end{align}
Note that the $E\left[\ldots\right]$ term on the righthand side of \eqref{E2} is $\text{Var}(\X|\bm{\beta}_1^\top\X)$ which, from Eq. 4 of \cite{PR2007}, is equal to $\Si-\Si\bbeta_1(\bbeta_1^\top\Si\bbeta_1)^{-1}\bbeta_1^\top\Si$.  Therefore, \eqref{E2} simplifies to just $E(R_h^2|Y\in S_h)(\bm{I}-\bm{P})\Si^{-1}(\bm{I}-\bm{P})$ which completes the proof.

\end{appendices}

\bibliographystyle{kbib}
\bibliography{references}
  
\end{document}